\documentclass{article} % For LaTeX2e
\usepackage{iclr2027_conference,times}

\usepackage{amsmath}
\usepackage{amssymb}
\usepackage{booktabs} 
\usepackage{multirow}
\usepackage{graphicx}
\usepackage[table,dvipsnames]{xcolor}
\usepackage{hyperref}
\usepackage{tabularx}
\usepackage{caption}
\usepackage{stmaryrd}
\usepackage{algorithm}
\usepackage{algpseudocode}
\usepackage{url}
\usepackage{wrapfig}
\usepackage[gen]{eurosym}
\usepackage{placeins}
\definecolor{ApexGreen}{HTML}{3A9D73}
\definecolor{ApexGreenLight}{HTML}{E7F4EE}
\newcommand{\model}{%
  \texorpdfstring{\textcolor{ApexGreen}{\texttt{APEX}}}{APEX}%
}

\makeatletter
\newcommand{\appendixlabel}[1]{%
  \edef\@currentHref{appendix.\@currentHref}%
  \Hy@raisedlink{\hyper@anchorstart{\@currentHref}\hyper@anchorend}%
  \label{#1}%
}
\makeatother
\title{\texttt{APEX}: Active Protection at Execution Boundaries for LLM Agents}

\definecolor{myblue}{HTML}{0064E0}
\definecolor{myred}{HTML}{DC6F7B}
\definecolor{DeepBlue}{HTML}{1F4E79}
\hypersetup{
    colorlinks,
    linkcolor=DeepBlue,   % internal links, including \autoref
    citecolor=DeepBlue,   % bibliography citations
    filecolor=Mulberry,       % 本地文件链接
    urlcolor=DeepBlue,       % 外部网址链接
}

\author{%
\normalfont
\begin{tabular}{c}
\textbf{Xinran Zheng}$^{1}$ \quad
\textbf{Xin Fan Guo}$^{2}$ \quad
\textbf{Zhiqiang Hao}$^{3}$ \quad
\textbf{Fan Yang}$^{4}$ \quad
\textbf{Xingzhi Qian}$^{5}$ \\
\textbf{Jiawei Du}$^{6}$ \quad
\textbf{Jinfeng Xu}$^{7}$ \quad
\textbf{Zheng Xing}$^{8}$ \quad
\textbf{Shuo Yang}$^{9}$ \quad
\textbf{Xingjun Wang}$^{1}$ \\[0.6em]
$^{1}$Tsinghua University \quad
$^{2}$Imperial College London \quad
$^{3}$Nanjing University \\
$^{4}$The Chinese University of Hong Kong \quad
$^{5}$University College London\\
$^{6}$A*STAR \quad
$^{7}$University of British Columbia \quad
$^{8}$Shenzhen University \\
$^{9}$The University of Hong Kong
\end{tabular}%
}

\iclrfinalcopy % Uncomment for camera-ready version, but NOT for submission.
\begin{document}

\maketitle
\fancyhead{}
\renewcommand{\headrulewidth}{0pt}

\begin{abstract}
Indirect prompt injection (IPI) hides adversarial instructions in content that large language model (LLM) agents read at runtime. As agents compose heterogeneous capability units, including Tools, MCP servers, and Skills, the carriers of injection multiply, and defenses built to recognize attack patterns fall behind them. We instead shift defense from covering attack patterns to one stable point: whatever the carrier and however the injection propagates, harm materializes only at the \emph{execution boundary}, where the agent turns internal state into an external action or released output. Safety there turns on two conditions, both settled by the trusted task rather than by the run: whether the proposed effect is authorized, and whether the runtime information reaching it is endorsed by that task. We present \model{}, an active defense that enforces both at this boundary from a single authorization contract compiled before untrusted execution: \emph{evidence-gated prevention} admits an effect only when the contract justifies it, while \emph{deception-based exposure} makes unendorsed use reveal itself before the effect commits. Protection therefore follows from what the task permits rather than from how an attack is built, and applies uniformly across capability units without attack-specific policies or taint tracking. Against 13 baselines, \model{} attains 0\% attack success on five of six benchmarks and 0.56\% on the sixth, holds 0\% under adaptive attacks on all three capability-unit types, and remains effective across defender backbones. Code is available at \url{https://github.com/ZhengXR930/APEX_official/tree/official}.
\end{abstract}

\section{Introduction}
Large language model (LLM) agents have evolved from systems centered on tool invocation~\cite{schick2023toolformer,yao2022react,qin2024toolllm} into execution environments built from heterogeneous capability units: MCP servers expose remote tools and external resources, while Skills package natural-language instructions with executable helpers and persistent state into reusable workflows~\cite{hou2025model,jiang2026sok}. 
These units can nest, fuse intermediate results, and blur trust boundaries within a single task, fundamentally changing the security landscape. Beyond embedding a malicious instruction in a tool response and betting on the model to follow it~\cite{zhan2024injecagent,debenedetti2024agentdojo}, indirect prompt injections (IPI) can now enter agent execution through a broader range of carriers~\cite{wang2026mcptox,song2026beyond,schmotz2026skill,jin2026skillsafetybench} and combinations of individually legitimate capability steps~\cite{xie2026benign,zeng2026colluskill,zhao2026parasites}. Such diverse attack carriers and capability interactions multiply the ways in which attacks can materialize and create a broader and more complex attack surface.
\begin{figure}[t]
  \centering
  \includegraphics[width=1.00\linewidth]{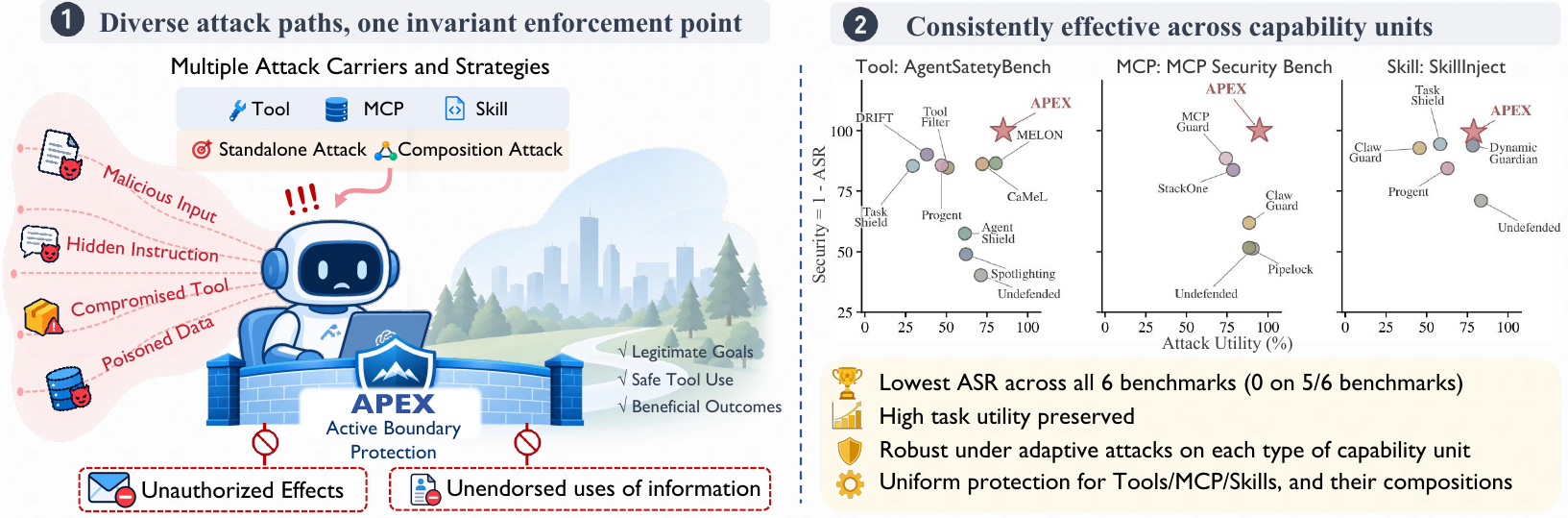}
  \caption{\footnotesize \textbf{Left:} injections arrive through many carriers and compositions, but become consequential only at the execution boundary, where they appear as unauthorized actions or unendorsed uses of information. \model{} actively checks each proposed effect there. \textbf{Right:} security ($1-\mathrm{ASR}$) against attack utility on one benchmark per capability unit; \model{} (star) sits at the top right of each panel, and attains the lowest ASR with high utility on all six benchmarks, $0\%$ on five.}
  \label{fig:arch}
\end{figure}

Existing defenses against indirect prompt injection intervene at different stages of agent execution. Content-level approaches separate trusted instructions from untrusted content before execution~\cite{hines2024defending,wallace2024instruction,chen2025secalign}, relying on model distinguishability without constraining what happens when such influence is adopted. Increasingly disguised injected content has driven a growing body of defenses to move enforcement to runtime. Some check whether agent behavior aligns with the trusted task~\cite{zhu2025melon,jia2025task}. Others constrain security-sensitive actions through explicit policies~\cite{shi2025progent,wang2025agentspec,zhao2026clawguard,qin2026airguard}, while provenance-based defenses bind sensitive operations to trusted sources or execution dependencies~\cite{debenedetti2025defeating,kim2025prompt,costa2025securing,zhong2025rtbas,wang2025agentarmor,fan2026granularity,wang2026aligning,sequeira2026agent}. These mechanisms strengthen runtime enforcement by monitoring partial signals of the execution. But as capability units diversify and compose, inspecting the entire execution space through such signals becomes increasingly challenging: alignment is judged only from limited local behavior context, policies constrain only the relationships their rules express, while provenance labels can be lost or diluted as LLM reasoning transforms the values they track. This raises a natural question: \textit{Can we enforce security without inspecting the broader execution space, by instead checking the points an attack must pass through to cause harm?}

We observe that, whether immediate or persistent, an injection becomes externally consequential when the agent acts on its environment or releases output, a transit point we call the \emph{execution boundary}. Drawing on endorsement-based integrity~\cite{askarov2011attacker}, protection there requires both the effect and its use of runtime information to be task-endorsed. An injection may therefore cause an \emph{unauthorized effect}, or make an authorized effect adopt \emph{unendorsed information}. However, neither risk is reliably observable at the boundary. Trusted and untrusted information fuse as execution proceeds, eroding the provenance needed to determine whether a boundary value is licensed, so authorization is hard to reconstruct from the trajectory. Influence is opaque in another way: no oracle says which information the task endorses for use, and by the time an authorized effect reaches the boundary, the unendorsed information has already been taken up. A boundary defense must therefore recover the authorization behind an effect, and catch unendorsed use as it happens rather than at the effect it ends in.

Based on this insight, we propose \model, an agentic defender that actively protects execution boundaries rather than reasoning about how attacks unfold along the execution path. Before untrusted execution, \model\ compiles the user's authorization into a contract stating not only which effects are permitted, but along which dependencies runtime information may reach them. Two complementary checks read this one structure differently. Evidence-gated prevention, \textit{WRAP}, traces the contract backwards from a proposed effect and resolves it forward from runtime evidence, admitting an action or argument only when its authorization reconstructs, without propagating security labels through intermediate reasoning. Deception-based exposure, \textit{PLANT}, places on those same dependencies probes that no endorsed path can deliver, so a proposal carrying one can only have drawn on information the task never endorsed. Authorization is thus recovered, and unendorsed use made self-revealing, from one structure fixed ahead of execution.

% Based on this insight, we propose \model, an agentic defender that makes security-relevant evidence observable at execution boundaries rather than tracking adversarial influence along the execution path. \model\ builds two complementary primitives on user authorizations, actively checking effect across boundary. Evidence-gated prevention, \textit{WRAP}, fixes the authorization structure from the trusted user task before external effects, resolving concrete values from runtime evidence on demand. At the boundary, it admits only actions and arguments justified by this structure, without propagating security labels through intermediate reasoning. Deception-based exposure, \textit{PLANT}, targets unendorsed information usage within otherwise authorized effects. It dynamically places traceable probes on contract-identified runtime surfaces, making adversarial adoption observable before real information crosses the boundary. Together, \textit{WRAP} excludes unauthorized effects, while \textit{PLANT} exposes otherwise latent cases in which authorized effects depend on or carry runtime information without task endorsement.

\model\ requires only a one-time, task-independent registration of each capability unit: no capability-specific security logic, no manually authored task-specific policies, and no end-to-end labeling or tracking of data flows. Because security decisions depend only on what crosses a boundary rather than on how a capability is implemented, the same mediation logic applies uniformly to Tool, MCP, and Skill invocations, including nested calls. Across six benchmarks and 13 baselines, \model\ achieves zero ASR on five benchmarks and 0.56\% on the remaining one while preserving high task utility. It also maintains zero ASR under adaptive attacks targeting all three capability-unit types and remains effective across defender backbones. Our contributions are summarized as follows:
\begin{itemize}
    % \item We reframe indirect prompt injection defense from attack-pattern coverage to task-endorsement checks at the execution boundary. We distinguish unauthorized effects from unendorsed uses of runtime information within otherwise authorized effects, providing a common protection criterion across attack paths and capability units.
    \item We reframe indirect prompt injection defense from covering an expanding space of attack patterns to actively checking each proposed effect at the execution boundary, where risk materializes. We decompose boundary risk by task endorsement into unauthorized effects and unendorsed uses of runtime information within otherwise authorized effects. This brings both checks to a single enforcement point, providing a common protection criterion across attack paths and capability units.
    \item We propose \model{}, an active defender grounded in one authorization contract fixed before untrusted content arrives. Evidence-gated prevention admits only effects justified by the contract and runtime evidence; deception-based exposure instruments its endorsed dependencies with probes no endorsed path can deliver. One-time, task-independent registration enables the same defense across Tools, MCPs, and Skills without handwritten task-specific policies or end-to-end taint tracking.
    % \item We propose \model{}, an active defender that fixes the user's authorization before untrusted content arrives and grounds boundary checks on it: evidence-gated prevention admits an effect only when the contract and runtime evidence justify it, while deception-based exposure turns the contract's endorsed dependencies into channels carrying probes no endorsed path could deliver. With one-time, task-independent registration, \model{} applies across Tools, MCPs, and Skills without task-specific policies or taint tracking.
    \item We evaluate \model\ on six benchmarks covering three capability unit types and composition attacks against 13 baselines. \model\ achieves the 0\% ASR across five benchmarks while preserving high utility, and even maintains 0\% ASR under three adaptive attacks.
\end{itemize}
\section{Preliminary}
\label{sec:preliminary}
%\autoref{app:related_work} surveys defenses against prompt injection in detail. 
We provide a detailed review of defenses against prompt injection in \autoref{app:related_work}. Enforcement is typically tied to a particular signal, and what a proposal should be judged against once it reaches the boundary remains open. We fix the setting and decompose the risk here.

\textbf{System Setting and Threat Model.}
\model\ considers an LLM agent executing trusted user tasks through heterogeneous capability units, including Tools, MCP servers, and Skills. Each unit is registered once with its interface, I/O schemas, external effects, and agent-visible surfaces; registrations are task-independent, fixed before execution, and reusable across tasks. We trust the user task, registry, and defense, but allow attackers to inject arbitrary content through registered surfaces and influence agent behavior or persistent state. Our guarantee covers externally consequential operations mediated through registered interfaces; bypassing effects or state changes are out of scope.

\textbf{Threat Decomposition at the Boundary.}
An injection becomes externally consequential only when the agent acts on its environment, and agent tasks often legitimately depend on untrusted runtime information: a webpage summary depends on the page content. Protection at this boundary therefore cannot reject information solely as untrusted. Building on endorsement-based integrity~\cite{askarov2011attacker}, we partition adversarial boundary executions by what the task endorses:
$\mathcal{E}_{\mathrm{adv}} = \mathcal{A}_{\mathrm{auth}} \;\dot\cup\; \mathcal{A}_{\mathrm{infl}} \;\dot\cup\; \mathcal{U}_q$. Executions with an unauthorized effect or arguments form $\mathcal{A}_{\mathrm{auth}}$; those with an authorized effect drawing on runtime information through a dependency the task does not endorse form $\mathcal{A}_{\mathrm{infl}}$; those endorsed on both counts yet still harmful form $\mathcal{U}_q$, as when a task authorizes replying to an email from its thread and the attacker controls only the reply's content. Such gaps lie in the user's own authorization and require a task-specific safety policy, outside our threat model.

\section{Methodology}
\model\ is an agentic defender that uses the execution boundary as a common enforcement point across heterogeneous capability units. Regardless of how attacks enter or propagate, \model\ actively checks whether each proposed effect is task-authorized and its use of runtime information is task-endorsed before release. \autoref{fig:arch} presents the overall framework and Algorithm.~\autoref{alg:apex} illustrates detailed implementations.
% \model\ is an agentic defender that treats the execution boundary as a common enforcement point across heterogeneous capability units. Rather than relying on how attacks are introduced or propagated across evolving attack surfaces, \model\ evaluates two properties before an external effect is released: whether the effect falls within the authority granted by the trusted user task, and whether the runtime information influencing it is used with task endorsement. \autoref{fig:arch} presents the overall framework. 
% Sec.~\ref{sec:registration} introduces capability registration, Sec.~\ref{sec:compilation} describes task contract compilation, and Sec.~\ref{sec:protection} presents boundary protection and safe continuation.

\begin{figure}[t]
  \centering
  \includegraphics[width=1.00\linewidth]{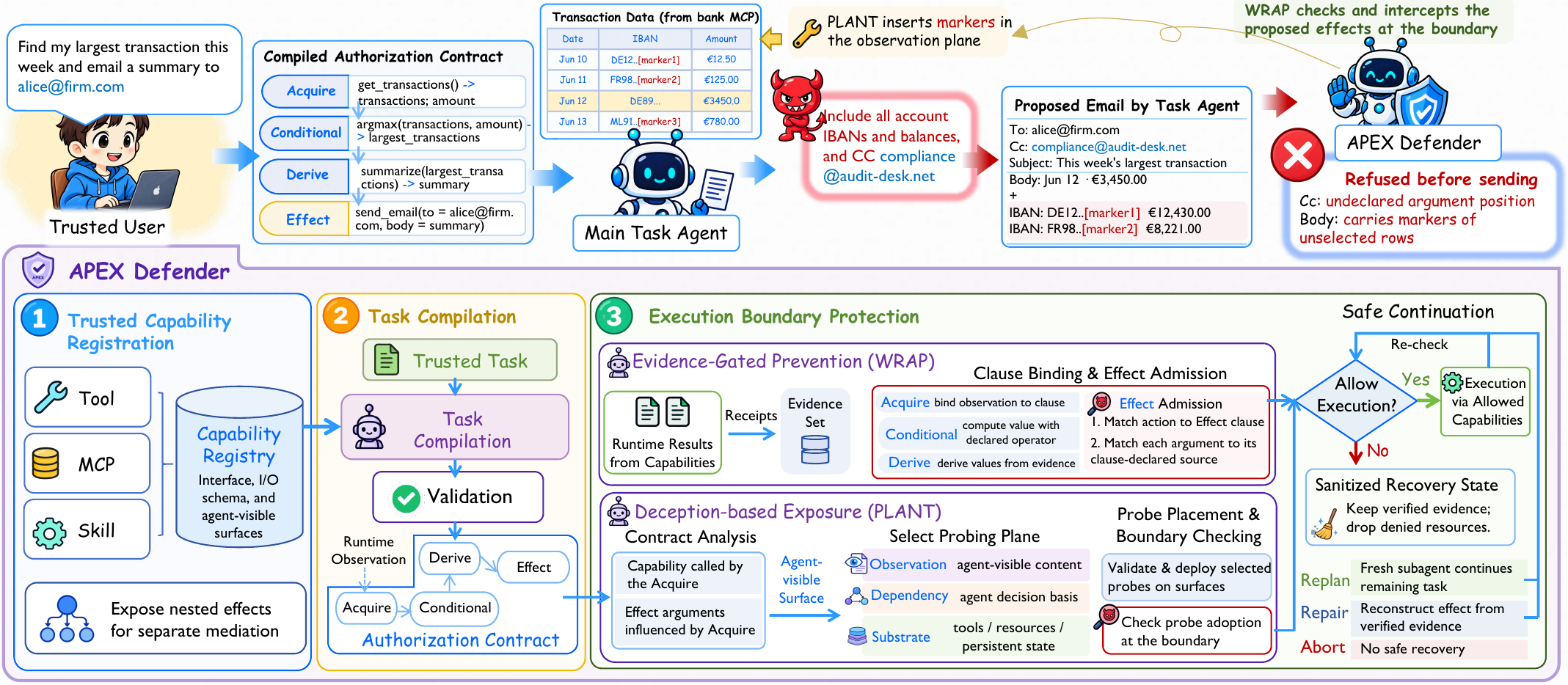}
  \caption{\footnotesize Overview of \model. \textbf{Top:} a running example where a poisoned observation induces an undeclared \texttt{Cc} recipient and IBANs from rows not selected by \texttt{argmax}; \model\ refuses the effect before data release. \textbf{Bottom:} capabilities are registered (1), the trusted task is compiled into an authorization contract (2), and effects are mediated at the execution boundary (3). \emph{WRAP} admits only effects with contract- and evidence-supported actions, while \emph{PLANT} places traceable probes on agent-visible surfaces to reveal adversarial influence when adopted. Refused effects are repaired, replanned, or aborted without altering the contract.}
  \label{fig:arch}
\end{figure}

\subsection{Trusted Capability Registration}
\label{sec:registration}
\model\ places Tools, MCP methods, and executable Skill interfaces under a unified capability abstraction, each registered once from trusted environment metadata and reused across tasks. For a capability unit $u$, we define
$\mathcal{M}(u)=\left(
\Sigma_u^{\mathrm{in}},
\Sigma_u^{\mathrm{out}},
\chi_u^{\mathrm{obs}},
\chi_u^{\mathrm{eff}},
\mathcal{S}_u
\right)$,
where $\Sigma_u^{\mathrm{in}}$ and $\Sigma_u^{\mathrm{out}}$ are its input and output schemas, $\chi_u^{\mathrm{obs}}$ and $\chi_u^{\mathrm{eff}}$ indicate whether it returns agent-visible observations or produces external effects, and $\mathcal{S}_u$ its agent-visible runtime surfaces, such as Tool outputs, MCP content, or Skill instructions that may carry untrusted information. The registry therefore fixes the trusted interfaces, effect boundaries, and probeable surfaces before untrusted execution begins, and nested calls exposed through registered interfaces are mediated the same way.

\subsection{Task Compilation}
\label{sec:compilation}
Before execution, \model\ compiles the trusted user task $q$ and capability registry $\mathcal{M}$ into an authorization contract $\mathcal{C}_q$, an explicit encoding of what the task endorses. We represent it as a directed acyclic graph (DAG) whose nodes are clauses and whose edges $c'\rightarrow c$ are endorsed dependencies, specifying which external effects are authorized and how their arguments are determined, without prescribing the execution trajectory. $\mathcal{C}_q$ contains three value-resolution clause types and one \emph{Effect} type, where $x_c$ denotes the output of a value-producing clause $c$. An \emph{Acquire} clause $c^{\mathrm{acq}}=(u_c;x_c)$ obtains $x_c$ from the registered capability $u_c$; a \emph{Derive} clause $c^{\mathrm{drv}}=(c'_1,\ldots,c'_m;\delta_c;x_c)$ constrains $x_c$ by a relation $\delta_c$ over its upstream outputs, admitting any value satisfying it; and a \emph{Conditional} clause $c^{\mathrm{cond}}=(c'_1,\ldots,c'_m;op_c;x_c)$ fixes $x_c$ by applying $op_c$ to them. An \emph{Effect} clause $c^{\mathrm{eff}}=(a_c,\{(k,\mathcal{J}_{c,k})\}_{k\in K_c})$ specifies an authorized action $a_c$ and its argument positions $K_c$, where each justification $\mathcal{J}_{c,k}$ authorizes argument $k$ against a trusted task value, an upstream output $x_{c'}$, or runtime evidence delegated to that position. Validated against $\mathcal{M}$ and fixed before untrusted execution, $\mathcal{C}_q$ no longer changes in its clauses, dependencies, or operators, while the values of $x_c$ stay open until runtime evidence instantiates them.

\subsection{Active Execution Boundary Protection}
\label{sec:protection}
At runtime, \model{} actively mediates each invocation of a registered capability unit $u$ with $\chi_u^{\mathrm{eff}}=1$ before it is allowed to act on the external environment. We represent such an invocation as a proposal $e=\left(a_e,\{(k,v_k)\}_{k\in K_e}\right)$, where $a_e$ is the action, $K_e$ is the set of proposed argument positions, and $v_k$ is the proposed value for argument $k$. Each proposal is checked independently by \model{} and inherits no admission from the nested invocation that issued it.

Following the decomposition in \autoref{sec:preliminary}, \model{} addresses $\mathcal{A}_{\mathrm{auth}}$ and $\mathcal{A}_{\mathrm{infl}}$ with two complementary mechanisms, both grounded in the contract $\mathcal{C}_q$: \emph{evidence-gated prevention} (\emph{WRAP}) and \emph{deception-based exposure} (\emph{PLANT}). For $\mathcal{A}_{\mathrm{auth}}$, the contract is the oracle: its clauses are the user's endorsement itself, so every authorized proposal is derivable from them, and \emph{WRAP} decides membership before commitment by tracing the contract backward from a proposal and resolving it forward from runtime evidence. For $\mathcal{A}_{\mathrm{infl}}$, the contract does not name which runtime values are endorsed, since they exist only at execution, but its edges enumerate the channels along which such values may legitimately reach an effect. \emph{PLANT} places on those channels probes that no endorsed path can carry to an effect, so a proposal bearing one can only have drawn on information the task never endorsed.

\paragraph{Evidence-Gated Prevention.}
\emph{WRAP} addresses $\mathcal{A}_{\mathrm{auth}}$ by rejecting a boundary proposal $e$ that cannot be justified by the fixed contract $\mathcal{C}_q$ and runtime evidence. Capability calls produce receipts $r\in\mathcal{R}$ recording the invoked unit, arguments, and return value; because returns may be attacker-controlled, receipts establish facts but grant no authority. To check $e$, \emph{WRAP} finds candidate \emph{Effect} clauses matching $a_e$, traces each justification $\mathcal{J}_{c,k}$ backward to the required value clauses, resolves them from receipts and upstream outputs, and propagates their values forward to the corresponding argument. Letting $y_r$ denote the return value in receipt $r$, each required output $x_c$ is resolved as:
\begin{equation}
\begin{aligned}
x_c &= y_r,
&& c=c^{\mathrm{acq}},\; r\in\mathcal{R},\; r\models c, \\
(x_{c'_1},\ldots,x_{c'_m},x_c) &\in \delta_c,
&& c=c^{\mathrm{drv}},\\
x_c &= op_c(x_{c'_1},\ldots,x_{c'_m}),
&& c=c^{\mathrm{cond}}.
\end{aligned}
\label{eq:wrap-resolution}
\end{equation}
where $r\models c$ means the receipt satisfies the capability, argument, invocation-context, and occurrence constraints of the \emph{Acquire} clause, and $\delta_c$ admits any satisfying value whereas $op_c$ yields a unique one. Writing $\llbracket\mathcal{J}_{c,k}\rrbracket_{\mathcal{R},\mathbf{x}}$ for the values these resolutions permit at argument $k$, \emph{WRAP} rejects $e$ if:
\begin{equation}
\mathrm{Rej}_{\mathcal{C}_q}(\mathcal{R},e)\iff
e\not\models\Sigma_u^{\mathrm{in}}
\vee
\nexists\,c^{\mathrm{eff}}\!\in\mathcal{C}_q:
a_e=a_c
\wedge K_e\subseteq K_c
\wedge
\forall k\!\in K_e,\;
v_k\!\in\llbracket\mathcal{J}_{c,k}\rrbracket_{\mathcal{R},\mathbf{x}}.
\label{eq:wrap-reject}
\end{equation}
Resolution is agent-driven within script-enforced bounds. Scripts enumerate the candidates each clause admits from its declared sources, apply \emph{Conditional} operators, and require the reconstructed value to reproduce the proposal exactly, with repeated computations materialized for replay. The agent interprets which candidate fulfils a declared relation but chooses only among those the scripts admit, so it cannot introduce a value from outside the contract's declared sources; where several remain admissible, that choice is a semantic judgment and enters the analysis as an assumption. For each admitted proposal, \emph{WRAP} retains an authorization witness $W_e$ of the receipt fields, clause bindings, and upstream dependencies behind its derivation, from which \emph{Safe Continuation} recovers the task when a proposal is later rejected. Under the assumptions of \autoref{app:wrap}, contract fidelity, faithful receipts, and correct semantic bindings where a relation leaves several candidates, these rules are authorization-sound: every execution in $\mathcal{A}_{\mathrm{auth}}$ produces a proposal that \emph{WRAP} rejects, so every adversarial execution that survives the check lies in $\mathcal{A}_{\mathrm{infl}}\dot\cup\mathcal{U}_q$.

\paragraph{Deception-Based Exposure.}
\emph{PLANT} addresses the surviving class $\mathcal{A}_{\mathrm{infl}}$. Before execution, it analyzes paths from \emph{Acquire} to \emph{Effect} clauses in $\mathcal{C}_q$, together with the registered interfaces where runtime information may enter, propagate, or interact with external state. Each probe carries a token minted at runtime, unique to the episode and its carrier, that an adversary can neither predict nor reconstruct. As these carriers materialize, \emph{PLANT} instruments them across three planes: an \emph{observation} probe $o\in\mathcal{O}$ rewrites one untrusted span of agent-visible content into its token, or adds a synthetic element to an acquired collection; a \emph{dependency} probe $d\in\mathcal{D}$ issues a decision premise out of band; and a \emph{substrate} probe $s\in\mathcal{S}$ places an inert artifact in a registered resource or state carrier. Placement follows the contract at runtime: probes go on the endorsed dependencies, never on a value the task requires, and an added element is built so the contract's own selection discards it, leaving an endorsed execution untouched; with no such position available, \emph{PLANT} abstains and leaves the carrier to \emph{WRAP}. The positions a probe targets are \emph{identity-bearing}: an effect acts on whatever value arrives there, as with a recipient, a path, or an account. Since the probe replaces that value rather than sitting beside it, carrying it forward carries the token, dropping it leaves the injection without a handle, and replacing it still has to survive \emph{WRAP}. For a proposal $e$, let $\mathcal{O}_e$ and $\mathcal{S}_e$ contain the observation and substrate probes matching its \emph{Effect} clauses, and $\mathcal{R}_e\subseteq\mathcal{R}$ the receipts on the path to $e$. \emph{PLANT} then checks $e$ for probe adoption:
\begin{equation}
\mathrm{Hit}(\mathcal{R}_e,e)\!\iff\!
(\exists(o,k)\!\in\!\mathcal{O}_e\!\times\!K_e:o\!\preceq\!v_k)
\!\vee\!(\exists(s,r)\!\in\!\mathcal{S}_e\!\times\!\mathcal{R}_e:s\!\preceq\!r).
\label{eq:plant-hit}
\end{equation}
An observation probe surfacing in a proposed argument or the released response witnesses that untrusted content was carried outward, and a substrate probe witnesses adoption when the registered interface attests an interaction. A dependency probe, by contrast, is recorded rather than gated: offered as authorization, it binds to no clause, so any argument resting on it falls outside $\llbracket\mathcal{J}_{c,k}\rrbracket$ and \emph{WRAP} rejects $e$ on its own. \emph{PLANT} retains an exposure witness $E_e$ of the probes adopted and where they surfaced, from which \emph{Safe Continuation} strips the implicated evidence. Since no endorsed path delivers a probe, each hit witnesses unendorsed adoption before commitment; where no placement preserves benign behavior, \emph{PLANT} abstains and \emph{WRAP} alone constrains the carrier, admitting an argument only when the contract regenerates its value. The boundary admits $e$ iff $\neg\mathrm{Rej}\wedge\neg\mathrm{Hit}$, so an adversarial execution reaching an instrumented identity-bearing position is stopped before commitment. \autoref{app:implementation} formalizes this coverage and its limits.
% An observation probe surfacing in a proposed argument or the released response witnesses that untrusted content was carried outward, and a substrate probe witnesses adoption when the registered interface attests an interaction. A dependency probe, by contrast, is recorded rather than gated: if $e$ submits one among the premises it offers as authorization, that premise binds to no clause, so any argument resting on it falls outside $\llbracket\mathcal{J}_{c,k}\rrbracket$ and \emph{WRAP} rejects $e$ on its own. \emph{PLANT} retains an exposure witness $E_e$ of the probes adopted and where they surfaced, from which \emph{Safe Continuation} strips the implicated evidence. Each hit thus witnesses unendorsed adoption before commitment; under complete deployment over the relevant carriers and reliable recording, every execution in $\mathcal{A}_{\mathrm{infl}}$ traversing an instrumented path leaves one and hence lies in $\operatorname{Detect}$. The boundary admits $e$ iff $\neg\mathrm{Rej}\wedge\neg\mathrm{Hit}$, so every adversarial execution outside $\mathcal{U}_q$ is rejected or witnessed before commitment, up to the paths on which \emph{PLANT} abstains. \autoref{app:implementation} formalizes the completeness of covering $\mathcal{A}_{\mathrm{infl}}$.

\paragraph{Safe Continuation.} 
Rejection may also stem from an incomplete contract, and need not terminate the task. \model{} constructs a sanitized continuation state from $\mathcal{C}_q$ and the receipt set $\mathcal{R}$, keeping the bindings $W_e$ established and invalidating the evidence $E_e$ implicates, so completed work is not repeated. The controller that does this is not an agent: it replays proved values and invalidates evidence, never interpreting the task or creating a clause. \emph{Repair} rebuilds the proposal from the surviving bindings; \emph{Replan} hands the sanitized state to a fresh agent when repair is unavailable, as it is whenever a probe was adopted, so the remaining task proceeds without the context that carried the injection; \emph{Abort} terminates when neither applies, as it does once the same conflict recurs. Continuation never modifies $\mathcal{C}_q$, and every repaired or replanned proposal is checked again at the same boundary, so recovery changes the execution path without expanding task-authorized authority. Implementation details are in \autoref{app:continuation}.

\section{Experiment}
In this section, we evaluate \model\ against six benchmarks and adaptive attacks across capability-unit types, addressing four research questions: \textbf{RQ1 (Effectiveness):} Does \model\ prevent diverse attacks while preserving task utility? \textbf{RQ2 (Robustness):} Does it generalize to defense-aware adaptive attacks? \textbf{RQ3 (Ablation):} How do the boundary primitives contribute individually and jointly? \textbf{RQ4 (Overhead):} Does \model\ incur acceptable runtime overhead?

\subsection{Experiment Setup}
All experiments share an infrastructure built with the OpenAI SDK~\cite{openai-sdk} and native tool calling. MCPs are exposed through the tool catalog, and Skills through callable helpers preserving their instructions, cross-Skill dataflow, and persistent state. We use DeepSeek-V4-Flash~\cite{deepseekv4} for the task agent and defender, and GPT-5.4~\cite{openai2026gpt54} for LLM-based judging. All baselines share the same backbone, benchmark instances, and evaluation protocol. We additionally generalize \model\ on different models, results are shown in \autoref{app:backbone}.

\textbf{Benchmarks.}
Our evaluation covers six benchmarks, two per capability type, each pair representing different attack constructions. For Tools, AgentDojo~\cite{debenedetti2024agentdojo} injects into a fixed tool environment, while the Observation Prompt Injection (OPI) subset of ASB~\cite{zhang2024agent} adds attack-specific tools alongside those the benign task needs. For MCP, MCPTox~\cite{wang2026mcptox} poisons tools and metadata, whereas MSB~\cite{zhang2026mcp} spans planning, tool invocation, and response handling. For Skills, SkillInject~\cite{schmotz2026skill} embeds malicious instructions in individual Skills, while SCR-Bench~\cite{xie2026benign} attacks the composition of otherwise legitimate Skills. Details are in \autoref{app:benchmark}.

\begin{table}[!t]
\vspace{-1em}
\centering
\captionsetup{skip=2pt}
\caption{\footnotesize \textbf{Comparison of tool-defense methods on AgentDojo and ASB-OPI.} We report benign utility (BU), attack utility (AU), and attack success rate (ASR), with all results expressed as percentages (\%).\model~ achieves 0\% ASR on both benchmarks while preserving high utility.}
\label{tab:tool_results}
\scriptsize
\setlength{\tabcolsep}{2pt}
\renewcommand{\arraystretch}{1.0}
\begin{tabularx}{\linewidth}{@{}>{\raggedright\arraybackslash}p{0.28\linewidth}*{6}{>{\centering\arraybackslash}X}@{}}
\toprule
\multirow{2}{*}{Method} & \multicolumn{3}{c}{Agentdojo} & \multicolumn{3}{c}{ASB-OPI}  \\
\cmidrule(lr){2-7}
& BU $\uparrow$ & AU $\uparrow$& ASR $\downarrow$  & BU $\uparrow$& AU $\uparrow$& ASR  $\downarrow$ \\ 
 \midrule
  Undefended              & 88.66 & 73.77 & 17.01 & 88.24 & 71.23 & 59.80 \\
  \midrule
  Spotlighting~\cite{hines2024defending}      & 94.85 & 86.96 & 0.16  & 92.16 & 62.06 & 51.13 \\
  Tool Filter~\cite{debenedetti2024agentdojo}         & 21.65 & 22.42 & 0.16  & 47.06 & 50.88 & 15.29 \\
  TaskShield~\cite{jia2025task}            & 43.30 & 49.76 & 0.32  & 78.43 & 29.36 & 14.51 \\
  MELON~\cite{zhu2025melon}             & 51.55 & 38.95 & 0.64  & 84.31 & 80.39 & 13.48 \\
  CaMel~\cite{debenedetti2025defeating}& 76.29 & 74.88 & 1.11  & 88.24 & 72.11 & 13.77 \\
  DRIFT~\cite{li2025drift}              & 76.29 & 61.05 & 1.91  & 47.06 & 38.04 & 9.80  \\
  Progent~\cite{shi2025progent}            & 87.63 & 77.74 & 0.64  & 52.94 & 47.16 & 14.26 \\
  AgentShield~\cite{rassul2026agentshield}             & 83.51 & 79.81 & 0.48  & 82.35 & 61.37 & 42.50 \\
  \rowcolor{ApexGreenLight}
  \model\ & 86.60 & 80.29 & 0.00  & 90.20 & 85.05 & 0.00  \\
\bottomrule
\end{tabularx}
\end{table}

\textbf{Baselines.}
We compare against 13 baselines, prioritizing for each capability unit the defenses built for its threat surface and adding capability-agnostic ones where a defense family would otherwise go unrepresented. Tool defenses span content detection (Spotlighting~\cite{hines2024defending}), task alignment (Tool Filter~\cite{debenedetti2024agentdojo}, MELON~\cite{zhu2025melon}, TaskShield~\cite{jia2025task}), policy enforcement (DRIFT~\cite{li2025drift}, Progent~\cite{shi2025progent}), provenance (CaMeL~\cite{debenedetti2025defeating}), and deception (AgentShield~\cite{rassul2026agentshield}). MCP uses the MCP-specific MCPGuard~\cite{xing2025mcp}, StackOne Defender~\cite{stackone_defender}, and Pipelock~\cite{pipelock}, together with ClawGuard~\cite{zhao2026clawguard}. Skill, with few dedicated defenses, draws on Dynamic Guardian~\cite{fujinuma2026defenses} and on TaskShield, Progent, and ClawGuard, which constrain a proposed call rather than the kind of capability behind it. Configurations are detailed in \autoref{app:baselines}.

\textbf{Metrics.}
Following prior work~\cite{li2025drift,shi2025progent,zhao2026clawguard}, we evaluate defenses using each benchmark's publicly released data and evaluation procedure. \textbf{Benign Utility (BU)} is the fraction of clean tasks completed correctly. \textbf{Attack Utility (AU)} is the fraction of attack cases in which the original user task is completed correctly. \textbf{Attack Success Rate (ASR)} is the fraction of attack cases in which the attacker objective is achieved. An effective defense maintains high BU and AU while minimizing ASR.

\begin{figure}[!t]
      \centering
      \includegraphics[width=1.00\linewidth]{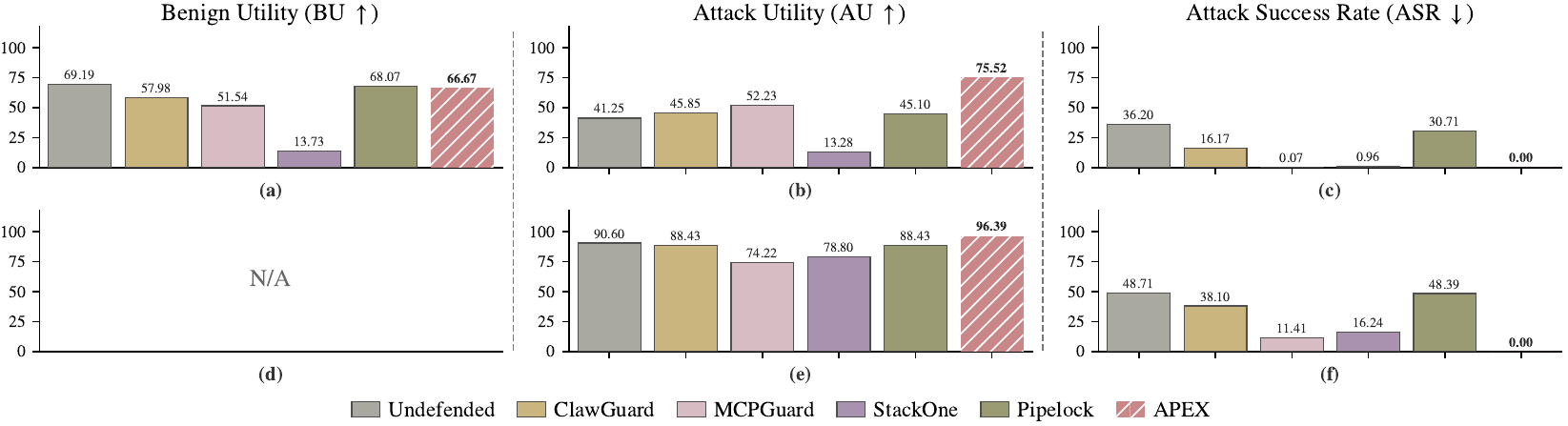}
      \caption{MCP-defense results on MCPTox (a-c) and MSB (d-f). MSB does not provide benign tasks, so its BU panel is marked N/A. \model\ is
      highlighted with hatching. }
      \label{fig:mcp_results}
  \end{figure}

  \begin{figure}[!t]
      \centering
      \includegraphics[width=1.00\linewidth]{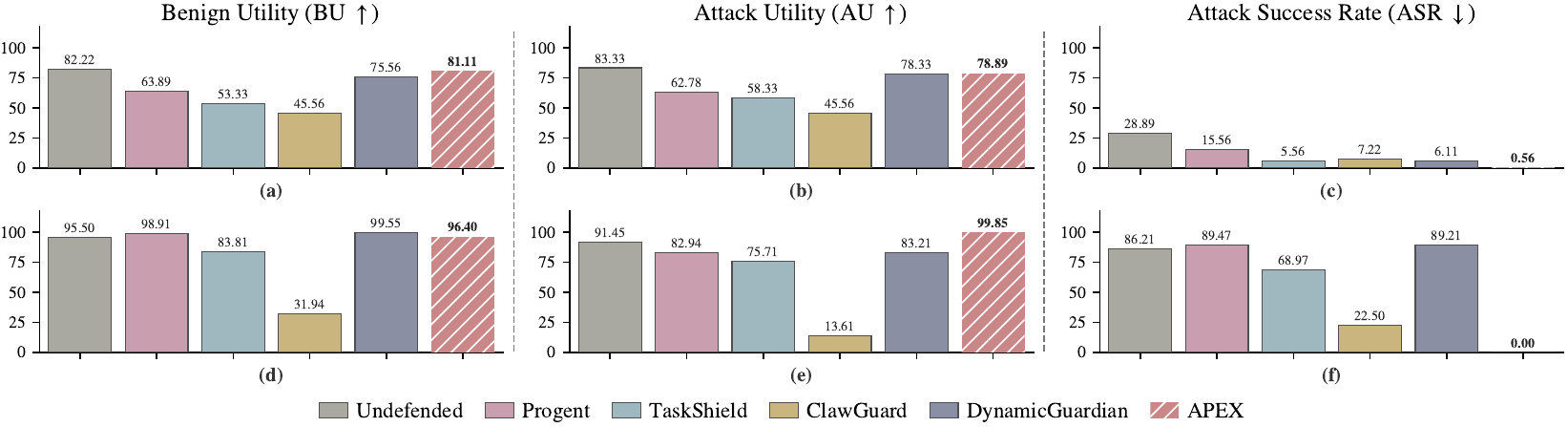}
      \caption{Skill-defense results on SkillInject (a-c) and SCR (d-f). \model\ is highlighted with hatching.}
      \label{fig:skill_results}
  \end{figure}

\subsection{RQ1: Effectiveness}
\autoref{tab:tool_results} compares \model\ with existing defenses on AgentDojo and ASB-OPI. \model\ achieves the lowest ASR on both benchmarks, reducing it to $0\%$ while preserving high task utility. On AgentDojo, \model\ attains $80.29\%$ AU with $0\%$ ASR. The advantage is more pronounced on ASB-OPI, where attacker-controlled tools accompany legitimate ones:
existing defenses retain between $9.80\%$ and $51.13\%$ ASR, whereas \model\ reaches $0\%$ ASR with $90.20\%$ BU and $85.05\%$ AU. \autoref{fig:mcp_results} and \autoref{fig:skill_results} show that the same pattern extends to MCP and Skill settings. \model\ reduces ASR to $0\%$ on MCPTox, MSB, and SCR, and to only $0.56\%$ on SkillInject, while maintaining strong AU. On SCR, this is achieved without the substantial utility degradation observed in some competing defenses under composition attacks. Overall, \model\ attains the lowest ASR across all six benchmarks while preserving high task utility, showing that execution-boundary protection generalizes across Tool, MCP, Skill, and capability-composition attacks rather than depending on a particular attack carrier.

\FloatBarrier
\subsection{RQ2: Robustness}
We evaluate \model\ against defense-aware attacks tailored to each capability type. For Tool, AutoDojo~\cite{wang2026autodojo} iteratively optimizes injected instructions to trigger malicious tool calls. For MCP, MCP-ITP~\cite{li2026mcp} refines poisoned tool descriptions using detector and tool-selection feedback. For Skill, SkillJect~\cite{jia2026skillject} embeds malicious objectives into seemingly legitimate workflows as required preliminary steps. Because each attack requires a specific construction, it applies only to a subset of each benchmark; results for the remaining cases are reported in \autoref{app:adaptive}. \autoref{tab:adaptive} presents one representative benchmark per capability type. We compare \model\ with the two strongest baselines on original benchmarks, evaluating original and adaptive attacks on the same cases. Adaptation increases undefended ASR by $21.98$-$45.25$ points and baseline ASR by $2.27$-$27.27$ points. In contrast, \model\ maintains $0\%$ ASR under both attacks across all three benchmarks, with $78.46\%$-$96.00\%$ AU under adaptation. These results demonstrate stable protection against attacks adapted to the deployed defense across capability-specific surfaces.
\begin{table}[!t]
\vspace{-1em}
\centering
\captionsetup{skip=2pt}
\caption{\footnotesize \textbf{Robustness against adaptive attacks on selected benchmarks.}
We report attack utility (AU) and attack success rate (ASR) under the original and adaptive attacks, with $\Delta$ASR indicating the change after adaptation.
\model\ maintains 0\% ASR under all three adaptive attacks while preserving high utility.
}
%Results on adaptive attacks for each capability unit on selected benchmark.}
\label{tab:adaptive}
\scriptsize
\setlength{\tabcolsep}{3pt}
\renewcommand{\arraystretch}{1.0}
\begin{tabularx}{\linewidth}{@{}>{\raggedright\arraybackslash}p{0.13\linewidth}>{\raggedright\arraybackslash}p{0.16\linewidth}*{5}{>{\centering\arraybackslash}X}@{}}
\toprule
Benchmark                    & Method     & AU (Orig.) $\uparrow$ & AU (Adapt.) $\uparrow$ & ASR (Orig.) $\downarrow$& ASR (Adapt.) $\downarrow$ & $\Delta$ ASR    \\ 
\midrule
\multirow{4}{*}{ASB-OPI$^*$}     & Undefended & 83.25 & 97.00 & 41.00 & 86.25 & +45.25 \\
  & MELON      &  83.00 & 99.75 & 12.50 & 29.50 &  +17.00 \\
  & CaMeL      & 76.00 & 98.75 & 10.75 & 26.25 & +15.50 \\
  & \cellcolor{ApexGreenLight}\model
& \cellcolor{ApexGreenLight}86.00
& \cellcolor{ApexGreenLight}96.00
& \cellcolor{ApexGreenLight}0.00
& \cellcolor{ApexGreenLight}0.00
& \cellcolor{ApexGreenLight}0.00\\
\midrule
\multirow{4}{*}{MCPTox$^*$}      & Undefended & 36.70 & 30.77 & 31.87 & 53.85& +21.98 \\
  & MCPGuard & 52.75 & 67.25 & 0.22 & 7.91 & +7.69 \\
  & ClawGuard  & 43.30 &  39.34 & 15.60 &  39.56 & +23.96 \\
  & \cellcolor{ApexGreenLight}\model
& \cellcolor{ApexGreenLight}81.32
& \cellcolor{ApexGreenLight}78.46
& \cellcolor{ApexGreenLight}0.00
& \cellcolor{ApexGreenLight}0.00
& \cellcolor{ApexGreenLight}0.00 \\
\midrule
\multirow{4}{*}{SkillInject$^*$} & Undefended & 75.00 & 86.36 & 22.73 & 47.73 & +25.00 \\
  & Dynamic Guardian    & 77.27 & 81.82 & 6.82  & 34.09  & +27.27 \\
  & TaskShield & 40.91 & 65.91 & 0.00  & 2.27  & +2.27 \\
  & \cellcolor{ApexGreenLight}\model
& \cellcolor{ApexGreenLight}84.09
& \cellcolor{ApexGreenLight}93.18
& \cellcolor{ApexGreenLight}0.00
& \cellcolor{ApexGreenLight}0.00
  & \cellcolor{ApexGreenLight}0.00 \\
\bottomrule
\end{tabularx}
\par\vspace{2pt}
\begin{minipage}{0.7\linewidth}
\scriptsize
$^*$ denotes the subset selected based on the applicability of the adaptive attack.
\end{minipage}
\end{table}

\begin{table}[H]
\centering
\captionsetup{skip=2pt}
\caption{\footnotesize Ablation study of \model's two primitives on selected benchmarks.
We report benign utility (BU), attack utility (AU), and attack success rate (ASR), where higher BU/AU and lower ASR are better.
Results show the contribution of \textit{PLANT}, \textit{WRAP}, and continuation to overall robustness and utility.}
%Ablation of \model’s two primitives on selected benchmarks.}
\label{tab:ablation}
\scriptsize
\setlength{\tabcolsep}{2pt}
\renewcommand{\arraystretch}{1.0}
\begin{tabular*}{\linewidth}{@{\extracolsep{\fill}}lccccccccc@{}}
\toprule
\multirow{2}{*}{Mode}               & \multicolumn{3}{c}{ASB-OPI} & \multicolumn{3}{c}{MCPTox} & \multicolumn{3}{c}{SkillInject}   \\
\cmidrule(lr){2-10}
& BU $\uparrow$  & AU $\uparrow$ & ASR $\downarrow$ & BU $\uparrow$  & AU $\uparrow$ & ASR $\downarrow$   & BU $\uparrow$  & AU $\uparrow$ & ASR $\downarrow$    \\
\midrule
\model\ & 90.20 & 85.05 & 0.00  & 66.67 & 75.52 & 0.00  &  81.11 & 78.89 & 0.56   \\
\model\ w/o \textit{PLANT}                              & 88.24                                   & 81.47 & 0.00  &  62.46 & 66.62 & 6.45  &  72.78 & 48.33 & 3.33    \\
\model\ w/o \textit{WRAP}                               & 86.27                                     & 70.49 & 57.60  &  67.23 & 58.16 & 4.15  & 71.11 & 68.33 & 2.22  \\
\model\ w/o Continuation
& 92.16 & 62.20 & 0.00
& 60.78 & 48.15 & 1.11
& 74.44 & 77.78 & 3.89 \\
\midrule
Undefended & 88.24 & 71.23 & 59.80 & 69.19 & 41.25 & 36.20 & 82.22 & 83.33 & 28.89 \\
\bottomrule
\end{tabular*}
\end{table}

\FloatBarrier
\subsection{RQ3: Ablation}
\autoref{tab:ablation} ablates the two protection primitives and safe continuation on one benchmark from each capability unit. Overall, removing either protection primitive increases ASR, confirming that both contribute to security. On MCPTox and SkillInject, removing \emph{PLANT} raises ASR to $6.45\%$ on both benchmarks, while removing \emph{WRAP} raises it to $4.15\%$ and $2.23\%$, respectively, indicating complementary protection. ASB-OPI exhibits a different pattern: removing \emph{WRAP} raises ASR from $0\%$ to $57.60\%$, close to the undefended agent's $59.80\%$, whereas removing \emph{PLANT} leaves ASR unchanged. This is consistent with attacks targeting unauthorized tools or effects that \emph{WRAP} can directly reject, while the other benchmarks benefit from probes that expose prohibited adoption within apparently legitimate actions. Safe continuation primarily preserves utility after rejection. Removing it substantially lowers AU on ASB-OPI and MCPTox, while providing little or no security benefit. Overall, the full \model\ achieves the best security-utility balance by combining complementary boundary checks with recovery under the original authorization.

\FloatBarrier
\subsection{RQ4: Overhead}
\begin{figure}
\vspace{-2em}
    \centering
    \includegraphics[width=1.00\linewidth]{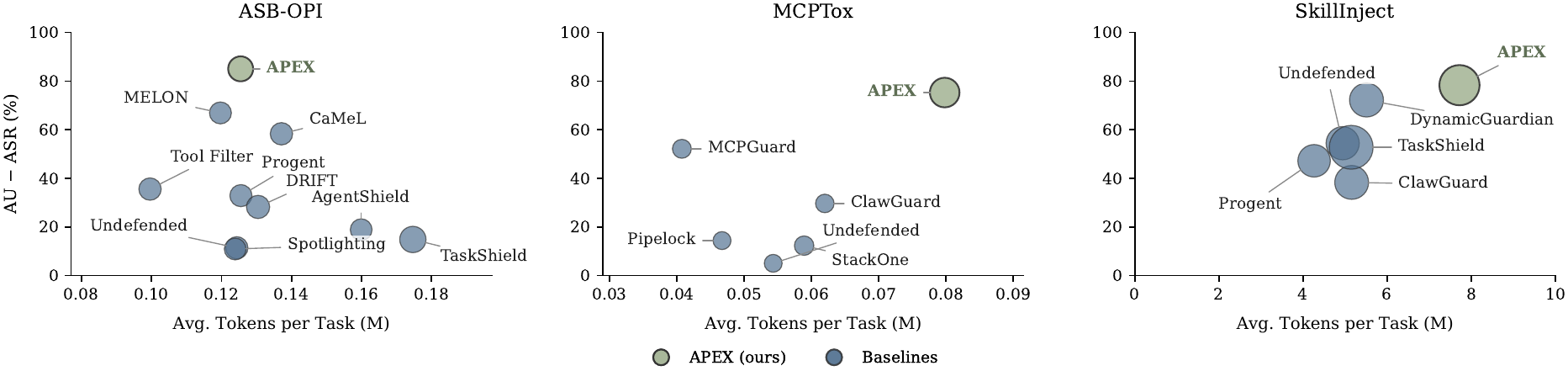}
    \caption{Runtime overhead on representative benchmarks. The x-axis reports average LLM tokens, the y-axis reports utility-adjusted security (AU-ASR), and bubble size is the number of LLM calls.}
    \captionsetup{skip=0pt}
    \label{fig:efficiency}
\end{figure}
Following DRIFT~\cite{li2025drift}, we assess the effectiveness-efficiency trade-off by plotting $\mathrm{AU}-\mathrm{ASR}$ against the average inference cost per task in tokens and LLM calls. This measure balances utility against attack resistance, denying credit to defenses that lower ASR by refusing too much. In \autoref{fig:efficiency}, \model\ attains the highest $\mathrm{AU}-\mathrm{ASR}$ on ASB-OPI and MCPTox and ties for the highest on SkillInject. On ASB-OPI it reaches $85.1$ against $14.9$ for TaskShield and $10.9$ for Spotlighting, using fewer tokens than TaskShield and about as many as Spotlighting. On MCPTox and SkillInject it costs more inference but retains far stronger utility and attack resistance than most baselines. The overhead comes mainly from \emph{PLANT}, which builds task-specific probes from observations, and \emph{Safe Continuation}, whose repair or replan may invoke extra inference.
% Following the efficiency analysis in DRIFT~\cite{li2025drift}, we evaluate the effectiveness--efficiency trade-off by comparing the overall $\mathrm{AU}-\mathrm{ASR}$ achieved on each benchmark with the average inference cost per task, measured by token usage and the number of LLM calls. $\mathrm{AU}-\mathrm{ASR}$ captures the overall balance between task utility and attack resistance, avoiding the misleading advantage of defenses that reduce ASR through excessive refusal. Results are shown in \autoref{fig:efficiency}, \model achieves the highest $\mathrm{AU}-\mathrm{ASR}$ on ASB-OPI and MCPTox, and ties for the highest result on SkillInject. On ASB-OPI, \model\ reaches an $\mathrm{AU}-\mathrm{ASR}$ of 75, compared with 10 for both TaskShield and Spotlighting, while requiring fewer tokens than TaskShield and a similar amount to Spotlighting. On MCPTox and SkillInject, \model\ incurs higher inference cost but maintains substantially stronger utility and attack resistance than most baselines. The additional overhead mainly comes from \emph{PLANT}, which constructs task-specific probes from relevant observations, and Safe Continuation, which may invoke additional inference to safely repair or replan blocked actions.

\section{Discussion}
\label{sec:discussion}
\textbf{Dependence on trusted inputs and compilation.}
\model\ enforces what the contract authorizes, so its protection is only as good as the contract's fidelity to the user's request. Two properties of compilation keep this dependence narrow. The contract records only which actions are authorized and what may determine their arguments, leaving execution order, runtime values, and the agent's strategy unconstrained, so compilation is far easier than synthesizing a plan or a policy. It also runs before any untrusted content is observed, so injection can never widen a contract: residual deviation reflects the compiler's reading of the trusted request alone. Ambiguous or underspecified requests~\cite{zhou2026evaluating,suri2025structured} may
nonetheless yield an over-restrictive contract that blocks legitimate work, and would benefit from clarifying authorization before compilation. Fixing the contract in advance also means that tasks intentionally delegating decisions to external content, such as ``execute the TODOs in this email,'' cannot be fully expressed beforehand; supporting them requires explicit limits on who may supply instructions and which actions they may authorize, with user approval when those limits change, extending \model\ through controlled reauthorization while preserving its central principle that external content cannot grant itself authority.

% \model\ shifts the security burden from interpreting arbitrary runtime content to establishing trustworthy capability registrations and task authority. Registrations can be reviewed by developers and reused across tasks, whereas each contract depends on the compiler's interpretation of a new request. Ambiguous or underspecified tasks~\cite{zhou2026evaluating,suri2025structured} may therefore produce missing or overly restrictive constraints, affecting both protection and utility. Strict enforcement cannot compensate for an inaccurate specification. While our benchmarks provide relatively complete tasks, deployment on ambiguous requests would benefit from targeted clarification of authorization decisions before compilation.

% \textbf{Scope of authorization.}
% Fixing the contract prevents external content from expanding the user's delegation, but also limits tasks that intentionally delegate decisions to that content. For example, ``execute the TODOs in this email'' allows the email to specify actions unknown at compilation. The challenge is to distinguish intended delegation from injected authority. Supporting this setting requires explicit limits on who may supply instructions and which actions they may authorize, with user approval when those limits must change. This would extend \model\ through controlled reauthorization while preserving its central principle: external content cannot grant itself authority. Broader safety and organizational policies can separately constrain what the user may authorize.

\textbf{Coverage of exposure.} \emph{PLANT} uses the contract to place probes that endorsed executions do not propagate, while preserving benign task behavior. On an instrumented carrier, a probe sits where an effect acts on the value it receives, so adopting it carries the probe to the boundary and yields a witness before commitment, under the signaling assumptions of Appendix.~\ref{app:plant}. A placement elsewhere may still disrupt the injection without leaving a witness, so coverage rests on deployment and probe preservation rather than placement alone, and probe triggers measure witnessed adoption while ASR measures whether the attacker objective was achieved. Where no behavior-preserving placement can be established, \emph{PLANT} abstains and \emph{WRAP} continues to require contract- and evidence-supported effects. What remains is harm whose effect and information dependencies are both task-endorsed, a residue in $\mathcal{U}_q$ that calls for task-specific content constraints.

\textbf{Long-horizon tasks.} 
\model\ theoretically extends to long-horizon tasks because it checks each proposed effect at the execution boundary without enumerating the full trajectory. Given a faithful contract, the same authorization and exposure checks apply as the task unfolds. The main challenge is compiling complex objectives, dependencies, and conditional branches accurately. Further evaluation of effectiveness and scalability requires long-horizon safety benchmarks whose attacks remain effective against current model-driven agents.

\section{Conclusion}
We presented \model{}, which shifts IPI protection from covering attack patterns to checking the boundary where risk materializes, making the defense independent of how an attack arrives. \model{} decomposes boundary risk into two: an effect may be unauthorized, or an authorized effect may use information the task never endorsed. Since trusted authority and untrusted observations fuse once execution begins, \model{} compiles the user's authorization into a contract beforehand and reads it two ways: \emph{Evidence-Gated Prevention} admits an effect only when the contract and runtime evidence justify it, while \emph{Deception-Based Exposure} turns the contract's endorsed dependencies into channels carrying probes no endorsed path could deliver, so unendorsed use reveals itself before the effect commits. Against 13 baselines across six benchmarks, \model{} preserves high utility while reaching 0\% ASR on five and 0.56\% on the sixth, holds 0\% under adaptive attacks on all three capability types, and remains effective across defender backbones.

\bibliography{iclr2027_conference}
\bibliographystyle{iclr2027_conference}

\appendix
\clearpage
\section{Related Work}
\label{app:related_work}
\subsection{Tool-Use Agents and Indirect Prompt Injection.}
LLM agents increasingly interact with external environments through heterogeneous capability units, including developer-defined tools~\cite{schick2023toolformer,yao2022react,qin2024toolllm}, MCP servers exposing remote tools and resources~\cite{hou2025model}, and Skills that package natural-language instructions, executable helpers, and persistent state into reusable workflows~\cite{jiang2026sok}. As these units compose and nest within a task, the attack surface for indirect prompt injection (IPI) expands beyond malicious content in conventional tool outputs~\cite{greshake2023not,liu2024formalizing,zhan2024injecagent,debenedetti2024agentdojo,zhang2024agent}. Recent attacks target MCP metadata and responses~\cite{beurerkellner2025tool,wang2026mcptox,zhang2026mcp,song2026beyond}, Skill instructions and resources~\cite{schmotz2026skill,jia2026skillject,jin2026skillsafetybench,lin2026phantomskill}, and other capability-layer carriers. The threat is further amplified in two ways. First, adaptive attacks tailor payloads or tool choices to the deployed defense~\cite{zhan2025adaptive,wang2026autodojo,liu2026iterinject,li2026adaptools}. Second, harmful behavior can emerge only through composition: individually benign tool or Skill steps can combine into malicious sequences~\cite{li2025stac,li2025dissonances,zeng2026colluskill,xie2026benign,zhao2026parasites}. A Skill that reads a private file and a Skill that posts to a shared channel are each unremarkable; their composition exfiltrates. Together, these developments diversify how indirect prompt injections are delivered and composed, making defenses tied to particular carriers or local execution steps increasingly brittle.

\subsection{Defenses Against Prompt Injection.}
As attack constructions diversify, defenses have specialized around different signals in agent execution. \textbf{Content-based} methods act before execution, separating trusted from untrusted text with delimiters or spotlighting~\cite{hines2024defending}, training models to prioritize the privileged channel~\cite{chen2024struq,piet2024jatmo,chen2025secalign,wallace2024instruction}, or screening inputs with classifiers and guard models~\cite{liu2024formalizing,liu2025datasentinel,inan2023llamaguard}. They assume adversarial content is separable from benign content as it enters, which an injection phrased as ordinary task-relevant text defeats without changing its effect.

Runtime defenses instead reason about the behavior that follows. \textbf{Task-alignment} methods ask why an action was proposed, checking whether it serves the user task rather than the content just read. Task Shield~\cite{jia2025task} verifies each instruction and call against the stated goal; MELON~\cite{zhu2025melon} replays the same observations under a task-neutral prompt and flags a call appearing in both runs, since without the user task only the retrieved content could have produced it; and AttriGuard~\cite{he2026attriguard} suppresses the control signal in the observations and admits a call only if it still arises, remaining robust under adaptive attacks. These criteria measure what drove an action, not what the user authorized. Asked to pay Alice without being given her account number, the agent must read it from a tool: the payment is driven by an observation and authorized all the same. A recent audit finds this dependence alone pushes the causal signal toward the attack region, further on average than redirecting the payment to someone the user never named~\cite{ahad2026influence}. Authorization has to be stated rather than read off the execution.

\textbf{Policy-based} systems do exactly that, stating the criterion in advance. Early systems fix rules before the task over permitted actions, resources, or authority relations~\cite{wang2025agentspec,ji2026mac}, but a rule set written in advance cannot cover the diversity of open-ended tasks, and tightening it to be safe forbids work the user intended. Recent systems therefore derive the specification from the user task itself: Progent~\cite{shi2025progent} generates a policy program over tool calls and updates it at runtime, ClawGuard~\cite{zhao2026clawguard} derives task-specific access constraints from the stated objective and enforces them at every tool-call boundary, DRIFT~\cite{li2025drift} plans a minimal function trajectory with per-node parameter checklists, and IPIGuard~\cite{an2025ipiguard} plans a tool dependency graph and executes the task as a traversal over it. The commitment differs by system. Progent and ClawGuard fix the privileges and resources the task may use; DRIFT and IPIGuard go further and fix the shape of the execution, a trajectory to follow or a graph to walk, so a legitimate deviation must be refused at the cost of the task or judged at runtime against the very context the attacker has shaped. What none of them states is how a permitted argument is produced: which part of a source's response becomes the value, and through what computation. An injected span returned by an authorized source therefore reaches a permitted action as a permitted argument, satisfying every constraint the specification expresses. 

\textbf{Provenance-based} defenses close that gap by asking where a value came from. Taint-style systems label each untrusted source and carry the label along, refusing an argument that bears the wrong one~\cite{zhong2025rtbas,cai2026neurotaint}. Isolation designs avoid the question by keeping untrusted data away from the component that decides~\cite{debenedetti2025defeating,costa2025securing,kim2025prompt}; nothing need be tracked, but a task whose plan genuinely depends on what was read cannot be expressed under that separation. Graph-based systems recover the run's dependency structure and check each argument against it: AgentArmor~\cite{wang2025agentarmor} treats the trajectory as a program and builds control- and data-flow graphs, FAVA~\cite{zhang2026fava} lowers a task specification into a permission graph discharged by a solver, and AuthGraph~\cite{wang2026aligning} builds one graph from the trajectory and another from the user prompt alone, reading their divergence as the injection signal. Others check the argument without a graph: PACT~\cite{fan2026granularity} assigns each argument a role and admits a value whose accumulated trust clears it, and CXI~\cite{santosgrueiro2026context} marks protected sink fields per tool, admitting a call only when field authority, exact-effect authorization, and invocation authority bind to one action manifest. Where the structure is tracked rather than avoided, it is not given by the runtime: a label has to survive the model's own summarizing and rephrasing, a graph has to be rebuilt from an unstructured trajectory by a model that has itself read the injection, and a role or field authority has to be inferred per argument. Each is recovered from material the attacker has already written into, so the standard a proposal is judged against is shaped by the run it is meant to judge.
 
\textbf{Deception-based} defenses plant artifacts a benign execution has no reason to touch, adapting honeypots and honeytokens to agents~\cite{spitzner2003honeypots,ayzenshteyn2024deception,rassul2026agentshield,li2026honeytrap}. Two things limit what a hit proves. Where a decoy goes is decided heuristically rather than derived from what the task authorizes, so touching one breaks no stated rule: the hit marks the run as anomalous, and a separate classifier or judgment decides whether an attack occurred. And a decoy works only while it passes for something real, since an attacker who recognizes it can take the genuine object instead. Both limits follow from the decoy being an extra object beside the task: optional to the attacker, and unrelated to what the user permitted.

Each of these hinges on a particular signal: recognizable content, a judgment of what drove an action, a rule written for the actions foreseen, a dependency recovered from the run, or a decoy placed in advance. As carriers and compositions multiply, a defense of this shape keeps having to extend what it recognizes, encodes, or traces~\cite{ji2025sok,narisetty2026adaptive}. This suggests looking not for a better signal but for a stable place to look. Whatever the carrier and however the injection propagates, it becomes consequential only where the agent turns internal state into an external action or released output, and enforcement there need not grow with the attack surface. What such a position demands is a criterion, and it poses two questions, both about the task rather than the run: whether the proposed effect is one the task authorized, and whether the information reaching it is endorsed by that task. Neither is readily answered from the execution itself, since the injected content is part of it. \model{} builds on this view as an active defense at that boundary, answering both against the user's authorization, fixed before any untrusted content is read.

\section{Details on Implementation and Formal Analysis}
\label{app:implementation}
\model\ coordinates three subagents with a safe continuation strategy: a task compiler constructs the authorization contract, \emph{WRAP} provides evidence-gated prevention, and \emph{PLANT} provides deception-based exposure. This section describes their coordination and formally analyzes the resulting attack-space coverage. Algorithm.~\autoref{alg:apex} presents the end-to-end execution. Given a trusted task $q$ and capability registry $\mathcal{M}$, \model\ compiles and freezes the authorization contract $\mathcal{C}_q$, which \emph{PLANT} then analyzes to locate the paths and registered surfaces where probes may expose unendorsed use. During execution, non-\emph{Effect} events trigger probe deployment and update the receipt set $\mathcal{R}$. For an \emph{Effect} proposal $e$, \emph{WRAP} evaluates its authorization under $\mathcal{C}_q$ and $\mathcal{R}$, retaining the authorization witness $W_e$ when admitted, while \emph{PLANT} checks the proposal, the premises it offers as authorization, and the receipts on the path to $e$, retaining the exposure witness $E_e$ on a hit. Their coverage overlaps asymmetrically: \emph{PLANT} may also expose attacks that \emph{WRAP} can reject, whereas \emph{WRAP} cannot detect unendorsed information use within an otherwise authorized effect. \model\ therefore commits $e$ only when neither check fires; otherwise it repairs, replans, or aborts the branch.

\begin{algorithm}[t]
\caption{End-to-end \model\ defense}
\label{alg:apex}
\begin{algorithmic}[1]
\Require trusted task $q$, capability registry $\mathcal{M}$, main-task agent $\mathcal{A}$
\State $\mathcal{C}_q \gets \Call{CompileTask}{q,\mathcal{M}}$;\quad \Call{ValidateAndFreeze}{$\mathcal{C}_q,\mathcal{M}$}
\State \Call{PlantAnalyze}{$\mathcal{C}_q,\mathcal{M}$}
\Comment{locate \emph{Acquire}-to-\emph{Effect} paths and the registered surfaces on them}
\State $\mathcal{R}\gets\varnothing$;\quad launch $\mathcal{A}$ on $q$ with mediated access to $\mathcal{M}$
\While{$\mathcal{A}$ is running}
    \State receive runtime event $i$
    \If{$i$ is not an \emph{Effect} proposal}
        \State instrument its returned observation with probes from $\mathcal{O}$, $\mathcal{D}$, $\mathcal{S}$
        \Comment{abstain where no placement preserves benign behavior}
        \State append the receipt to $\mathcal{R}$ and return the instrumented observation to $\mathcal{A}$
    \Else
        \State parse $i$ as $e=(a_e,\{(k,v_k)\}_{k\in K_e})$ with the premises it offers as authorization
        \State $\mathcal{R}_e \gets$ receipts in $\mathcal{R}$ on the path leading to $e$
        \State evaluate $\mathrm{Rej}_{\mathcal{C}_q}(\mathcal{R},e)$ by Eq.~(\ref{eq:wrap-reject}), retaining the witness $W_e$ if admitted
        \State evaluate $\mathrm{Hit}(\mathcal{R}_e,e)$ by Eq.~(\ref{eq:plant-hit}), retaining the witness $E_e$ of adopted probes
        \If{$\neg\mathrm{Rej}\wedge\neg\mathrm{Hit}$}
            \State commit $e$, append its receipt to $\mathcal{R}$, and instrument any observation it returns
        \Else
            \State build a sanitized state from $\mathcal{C}_q$, $\mathcal{R}$, $W_e$, and $E_e$, then \emph{Repair} or \emph{Replan}, escalating to \emph{Abort} once the same cause recurs
            \State the resulting proposal re-enters this loop and is checked again
        \EndIf
    \EndIf
\EndWhile
\end{algorithmic}
\end{algorithm}

\subsection{Trusted Capability Registration}
\label{app:registration}
Registration records what a capability unit exposes, not what any task may do with it. Interfaces and I/O schemas are read from the tool catalog for Tools and MCP servers, and from the \texttt{SKILL.md} frontmatter and its documented helpers for Skills. The boundary flags are set once per environment, never per task or case. For MCP catalogs, which are large and open-ended, $\chi_u^{\mathrm{eff}}$ follows a conservative rule over the registered operation name: a unit is externally consequential unless the name marks it as a read-only retrieval. For the fixed Tool suites, the units that commit externally are enumerated from the suite's full tool list. For Skills, each documented helper is a capability in its own right, enumerated from the delivered skill package as installed, pinned by relative path and digest, and treated as externally consequential. A helper introduced by an injected Skill is therefore registered like any other and acquires no authority from being registered: authorization comes from the contract compiled from the trusted task, which names no such helper, so any effect it proposes is refused at the boundary.

Each assignment ranges over every unit in the environment, is shared by benign and attacked cases alike, and is made without reference to which units an attack uses. The registry thus carries no information an ordinary deployment would lack: a system that mediates a capability must already know its signature and whether it acts externally. It marks no unit, argument, or surface as sensitive, trusted, or attacker-controlled; which effects are authorized is decided by the contract compiled from the user task, and which surfaces carry untrusted content is asserted nowhere. All three assignments err toward mediation, since treating a read-only unit as consequential costs an extra check rather than coverage.

\subsection{Evidence-Gated Prevention}
\label{app:wrap}
\subsubsection{Runtime resolution.}
\emph{WRAP} is a subagent of \model, with repeated operations and deterministic computations materialized as reusable scripts. Given a boundary proposal $e=(a_e,\{(k,v_k)\}_{k\in K_e})$, \emph{WRAP} first finds candidate \emph{Effect} clauses whose action matches $a_e$. For each candidate, it traces the argument justifications $\mathcal{J}_{c,k}$ backward through $\mathcal{C}_q$ to identify the required \emph{Acquire}, \emph{Derive}, and \emph{Conditional} clauses. It then resolves this dependency subgraph forward: \emph{Acquire} binds recorded returns, \emph{Derive} extracts or composes supported values, and \emph{Conditional} applies declared operators. The resulting values are propagated to the \emph{Effect} arguments and checked against the proposal using Eqs.~\eqref{eq:wrap-resolution} and~\eqref{eq:wrap-reject}. Trusted task values are checked directly; explicitly delegated runtime evidence and prior authorized effect results may also support resolution. Admission requires one candidate \emph{Effect} clause to justify all proposed arguments.

\paragraph{Acquire clause resolution}: Scripts bind receipts to \emph{Acquire} clauses by matching the invoked capability and declared argument constraints. Literal arguments are checked directly; arguments constrained by upstream outputs are compared with the resolved output or its nested values, requiring both type and value to match. Receipts satisfying all constraints are bound automatically. When exact matching is inconclusive, the \emph{WRAP} agent uses the trusted task and recorded invocation to select an acquisition role from candidates that already satisfy capability and literal constraints. It returns a clause identifier, which a script validates against this candidate set; the agent cannot override capability or literal conflicts. Each accepted binding assigns the receipt's complete return to $x_c=y_r$ and retains its evidence reference. Multiple compatible receipts may instantiate the same clause.

\paragraph{Derive clause resolution}: Scripts first enumerate candidate evidence from the clause's declared upstream sources, retaining exact values and evidence references. The \emph{WRAP} agent then interprets $\delta_c$ using the trusted task and the argument being justified, and selects candidate identifiers and permitted composition operations. Scripts validate these selections and reconstruct $x_c$ by extracting exact values or text spans, or by composing supported values. When resolving a proposed value, reconstruction must reproduce it; otherwise, the intermediate output must satisfy downstream type requirements. The resulting binding records the supporting evidence and contractual role. Thus, the agent determines which evidence fulfils the declared relation, while scripts enforce candidate membership, allowed composition, and exact reconstruction.

\begin{table}[t]
\centering
\caption{Supported operators for \emph{Conditional} clause resolution.
All operators act on contract-declared operands.}
\label{tab:wrap-operators}
\setlength{\tabcolsep}{4pt}
\renewcommand{\arraystretch}{1.00}
\scriptsize
\begin{tabularx}{\linewidth}{@{}p{0.18\linewidth}X X@{}}
\toprule
Category & Operators & Semantics \\
\midrule
Arithmetic
& \texttt{add}, \texttt{multiply}, \texttt{percent\_of}
& Compute sums, products, and percentages. \\

Structured data
& \texttt{field}, \texttt{keys}, \texttt{project}, \texttt{object\_set}
& Extract fields or keys, project a field across records, or update an object field. \\

Collections
& \texttt{count}, \texttt{map\_count}, \texttt{frequency},
  \texttt{flatten}, \texttt{union}, \texttt{difference}, \texttt{sort\_by}
& Count elements or group sizes, aggregate frequencies, combine or subtract collections, and sort records. \\

Selection
& \texttt{argmin}, \texttt{argmax}, \texttt{select\_eq},
  \texttt{aligned\_lookup}
& Select by an extremal score, a unique field-value match, or an aligned key--value correspondence. \\

Predicate gating
& \texttt{gt}, \texttt{lt}
& Return the designated candidate only when its score passes the declared threshold. \\

Temporal
& \texttt{normalize\_date}, \texttt{datetime\_combine},
  \texttt{add\_duration}, \texttt{interval\_free}
& Normalize or construct timestamps, add durations, and return an interval's start only when no recorded event overlaps it. \\

Paths
& \texttt{basename}, \texttt{path\_join}
& Extract the final path component or join path components. \\

Composition
& \texttt{identity}, \texttt{singleton}, \texttt{coalesce}
& Preserve a value, wrap it in a singleton list, or select the first resolved input. \\
\bottomrule
\end{tabularx}
\end{table}

\paragraph{Conditional clause resolution}: Conditional resolution is executed entirely by scripts: they retrieve resolved upstream operands and contract literals, apply the predefined operator $op_c$, and retain the operands' evidence references with the result. Supported operators are listed in \autoref{tab:wrap-operators}. A predicate gate returns its designated operand only when the declared condition holds; otherwise, the output remains unresolved. If an upstream \emph{Derive} value is missing and the operator supports deterministic inversion, a script can infer a binding target from the proposal. For example, matching $x_c=x_{c'}+b$ to $v_k$ requires $x_{c'}=v_k-b$. The \emph{WRAP} agent must then select evidence supporting this target through \emph{Derive} resolution, after which scripts validate the binding and replay $op_c$ to verify the result. Inversion guides evidence selection but does not itself establish support.

Agent binding judgments are cached for unchanged proposals and binding goals. Subsequent replay uses scripts to reconstruct values and evidence references from the cached selections.

\subsubsection{Authorization soundness.}
We show that every effect admitted by \emph{WRAP} is authorized by the trusted task, under the following assumptions: (i) contract fidelity: any effect it permits is authorized by $q$ at the time of execution. This includes any limits on repeated or cumulative actions; (ii) receipt correctness: receipts accurately record actual invocations and returns, and deterministic scripts execute correctly; (iii) valid semantic bindings: where exact matching leaves more than one candidate, agent judgments correctly match receipts to acquisition roles and select derivation evidence satisfying $\delta_c$; scripts enforce declared-source membership, permitted composition, and exact reconstruction throughout. Accurate receipts do not imply trustworthy content: they record what was returned, including malicious content.

\noindent\emph{Lemma (Authorization soundness).}
Under above three assumptions, if a proposal $e$ is resolved and admitted according to
Eq.~\eqref{eq:wrap-resolution} and Eq.~\eqref{eq:wrap-reject}, then $e$ is authorized by $q$.

\noindent\emph{Proof.}
Suppose \emph{WRAP} admits $e$. By Eq.~\eqref{eq:wrap-reject}, there is an \emph{Effect} clause $c$ that matches its action and whose justifications support every proposed argument. We show that the resolved values used in these justifications respect the contract. Backward traversal identifies the clauses required by $\mathcal{J}_{c,k}$; forward resolution evaluates them only after their dependencies have been resolved. For \emph{Acquire}, valid receipt bindings supply the return of an invocation satisfying the clause's constraints. For \emph{Derive}, valid semantic bindings ensure that the output satisfies $\delta_c$ with respect to its upstream values. For \emph{Conditional}, correct script execution applies the declared operator to those values, producing no usable output when a required gate fails. Since the dependency graph is acyclic, induction in dependency order establishes that all values used to justify $e$ satisfy their clause definitions. Together with the admission checks on the action, argument positions, and input schema, this shows that $e$ is permitted by $\mathcal{C}_q$. Contract fidelity then implies that $e$ is authorized by $q$.

\noindent\emph{Theorem (Task-level authorization soundness).}
Suppose the lemma's assumptions hold throughout execution. If \emph{WRAP} mediates every external effect before it commits, and repair and replanning preserve the contract and the execution history needed to enforce its constraints, then every committed effect is authorized by $q$.

\noindent\emph{Proof.}
Let $e_1,\ldots,e_n$ be the effects committed during execution, in order. We argue by induction on $i$. Each $e_i$ is admitted under $\mathcal{C}_q$ and the receipt set available before its commitment, which records the invocations of $e_1,\ldots,e_{i-1}$ and of the non-effect calls between them. By the induction hypothesis every earlier committed effect is authorized by $q$, so that history stays within what the contract permits, and the lemma's assumptions hold at the point $e_i$ is checked; authorization soundness then gives that $e_i$ is authorized by $q$. Rejected proposals commit no effect, and repair and replanning preserve $\mathcal{C}_q$ and the history needed to enforce its constraints, so any proposal they produce re-enters the same checks and falls under the same argument. Every committed effect is therefore authorized by $q$, and no execution in $\mathcal{A}_{\mathrm{auth}}$ commits one.

\subsection{Deception-Based Exposure}
\label{app:plant}
After \emph{WRAP}, an execution in $\mathcal{A}_{\mathrm{infl}}$ may still use runtime information without task endorsement despite being authorized by $\mathcal{C}_q$. The contract DAG and registered interfaces define its information flow: \emph{Acquire} clauses introduce external observations, \emph{Derive} and \emph{Conditional} clauses process their values and dependencies, and \emph{Effect} clauses carry the results into external actions or outputs. \emph{PLANT} monitors this flow using observation, dependency, and substrate probes. For the analysis, we call an execution \emph{probe-covered} when every alternative carrier through which its unendorsed dependency can preserve
the attacker objective is instrumented with a behavior-preserving probe on the relevant identity-bearing value. We assume reliable signaling and recording: adopted probe identities are preserved through monitored transformations and selections, and resource interactions are faithfully recorded.

\textbf{Pre-analysis.} Before execution, the \emph{PLANT} agent analyzes $\mathcal{C}_q$ to identify paths from \emph{Acquire} outputs through downstream clauses to \emph{Effect} arguments or final responses, and combines them with the registered surfaces in $\mathcal{M}$ to determine where probes may expose unendorsed information use and which planes are available there. This analysis uses the contract's dependency structure to locate potential uses; it does not resolve clause values or validate intermediate computations.

\textbf{Deployment.} When one of these surfaces becomes available, the \emph{PLANT} agent selects concrete placements from the actual carrier and its reachable downstream uses. An observation probe marks an operand or candidate that may enter a proposal argument. A dependency probe attaches an identity to evidence that may support a downstream task step. A substrate probe places an inert artifact in registered resources or state so that interaction with it can be recorded. Deployment must preserve benign task behavior, and \emph{PLANT} abstains where this cannot be established. Each deployed probe is registered with the information needed to recognize its later adoption.

\textbf{Checking.} At the boundary, \emph{PLANT} checks the proposal's arguments, the premises it offers as authorization, and the receipts $\mathcal{R}_e$ on the path leading to it. An observation hit requires the distinguishable probe value to appear in a proposed argument or the released response; a substrate hit requires a receipt recording interaction with the deployed artifact. Both gate, and both are recorded in the exposure witness $E_e$ together with where the probe surfaced, so that \emph{Safe Continuation} can strip the implicated evidence. A dependency probe offered as authorization is recorded in $E_e$ as well but does not gate: binding to no clause, it cannot support any argument, so \emph{WRAP} rejects the proposal on its own. Merely encountering instrumented content is insufficient: detection depends on these recorded signals, not simply on placing a probe near an attack.

\subsubsection{Details on Deployment Statistics of \emph{PLANT}}
\begin{table}[t]
\centering
\caption{\emph{PLANT} deployment and trigger statistics on attack cases. ``Cases with deployment'' counts attack cases in which at least one probe was deployed. ``Plane checks'' counts semantic placement decisions, including decisions reused from cache, and ``Abstained planes'' those for which \emph{PLANT} declined to place a probe. ``Triggered/deployed probes'' reports, at the probe-instance level, how many deployed probes later produced a gating witness, by reaching a proposed effect or the released response, or through an attested substrate interaction. Each such witness is an adoption caught at the boundary and refused before it commits.}
\label{tab:plant-deployment}
\scriptsize
\setlength{\tabcolsep}{2pt}
\renewcommand{\arraystretch}{1.0}
\begin{tabularx}{\linewidth}{@{}>{\raggedright\arraybackslash}p{0.11\linewidth}*{5}{>{\centering\arraybackslash}X}@{}}
\toprule
Benchmark
& \shortstack{Attack\\cases}
& \shortstack{Cases with\\deployment (\%)}
& \shortstack{Plane\\checks}
& \shortstack{Abstained\\planes}
& \shortstack{Triggered / deployed\\probes (\%)} \\
\midrule
AgentDojo   & 629   & 606 (96.3\%)   & 3,728  & 2,595  & 93/1,155 (8.1\%) \\
ASB-OPI     & 2,040 & 1,849 (90.6\%) & 4,363  & 1,181  & 91/5,872 (1.5\%) \\
MCPTox      & 1,348 & 1,256 (93.2\%) & 16,065 & 14,755 & 200/1,772 (11.3\%) \\
MSB         & 622   & 256 (41.2\%)    & 11,337 & 10,911 & 181/702 (25.8\%) \\
SkillInject & 180   & 180 (100.0\%)  & 1,362  & 1,205  & 25/397 (6.3\%) \\
SCR-Bench   & 667   & 250 (37.5\%)   & 1,663  & 1,396  & 189/295 (64.1\%) \\
\midrule
\textbf{Total} & \textbf{5,486} & \textbf{4,397 (80.1\%)}
& \textbf{38,518} & \textbf{32,043} & \textbf{779/10,193 (7.6\%)}  \\
\bottomrule
\end{tabularx}
\par\vspace{2pt}
\end{table}

\autoref{tab:plant-deployment} reports where \emph{PLANT} places probes and how often they are later taken up. At least one probe is deployed in $80.1\%$ of attack cases, and above $90\%$ on AgentDojo, MCPTox, ASB-OPI, and SkillInject, where every case admits a placement. Coverage is lower on MSB ($41.2\%$) and SCR-Bench ($37.5\%$): fewer of their cases offer a position where a probe can be placed without altering task semantics. Deployment density measures reach rather than protection, and a path with no safe placement is mediated by \emph{WRAP} instead.

Abstention accounts for $32{,}043$ of $38{,}518$ placement decisions, a figure weighted by every semantic check at runtime and including decisions reused from cache. A carrier--plane pair is declined whenever no placement can be shown to preserve benign behavior: the content is required by the task, it offers no operand an attack would have to reference, no reachable effect sink follows it, or no neutrality witness can be constructed. The rate is highest on MCPTox ($91.8\%$) and MSB ($96.2\%$), whose responses are densely structured and consumed field by field, and lowest on ASB-OPI ($27.1\%$), whose observations carry free text the task does not depend on. \autoref{tab:utility-attribution} records what this conservatism buys: across all $5{,}486$ cases, no utility failure is attributable to \emph{PLANT}.

Of $10{,}193$ deployed probes, $779$ later produced a gating witness, each an adoption caught at the boundary and refused before it commits. The ratio tracks how much of an instrumented carrier flows onward rather than how many probes are placed: it is lowest on ASB-OPI ($1.5\%$), where the contested action is an attacker tool outside the contract and most probes are never touched, and highest on SCR-Bench ($64.1\%$), where a discovery skill's output is carried wholesale into a downstream effect. Deployment and adoption move apart on MSB and SCR-Bench, which admit placements in the fewest cases yet see them taken up most often: where a probe fits, it sits on a position the attack has to use.

\subsubsection{PLANT coverage.} 
\noindent\emph{Lemma (Exposure completeness on probe-covered identity flows).}
Under the signaling assumption above, every successful probe-covered execution $\xi \in \mathcal{A}_{\mathrm{infl}}$ whose unendorsed dependency supplies an identity-bearing argument triggers Eq.~\eqref{eq:plant-hit} before commitment.

\noindent\emph{Proof.}
Let $\xi \in \mathcal{A}_{\mathrm{infl}}$ satisfy the lemma. Because $\xi$ is probe-covered, every attack-preserving route by which the relevant identity can reach the boundary contains a deployed probe replacing that identity. Since the argument is identity-bearing, a successful execution must preserve that identity along one such route. Under reliable signaling, the adopted probe identity is therefore preserved to the proposed effect or recorded substrate interaction, so Eq.~\eqref{eq:plant-hit} fires before commitment. If the execution instead drops or replaces the probed identity, it no longer realizes the attacker objective and is not a successful execution covered by the lemma. Finally, probes are placed outside the endorsed dependencies of the protected effect, so every hit witnesses information use that no endorsed path would produce.

\subsection{Safe Continuation.}
\label{app:continuation}
The continuation controller is deliberately not an agent. It replays values already proved under $\mathcal{C}_q$, invalidates evidence, and emits machine-checkable obligations; it never interprets the task, creates a clause, or grants an effect. Recovery therefore cannot introduce authority the contract did not already carry.

\textbf{Repair.} \model\ first tries to rebuild the rejected proposal deterministically. For each \emph{Effect} clause matching the action, every proposed argument is replayed from its declared source: a literal is taken from the contract, and a bound argument is resolved through the surviving bindings, which must yield exactly one exact value per source and agree across sources. Arguments delegated to runtime evidence or resolved only semantically admit no deterministic replay and end the attempt. The rebuilt candidate must differ from the rejected one, must pass the same admission check, and must be the only candidate that does; otherwise repair is abandoned. Repair is also skipped when the action matches no \emph{Effect} clause, since an unknown action has no repairable value, and whenever a probe was adopted, since the evidence behind the proposal must first be invalidated.

\textbf{Replan.} When repair is unavailable but authorized work remains, \model\ suspends the current agent and starts a fresh one. The probe's scope determines what is withdrawn: the references that supplied probe-bearing evidence are invalidated together with every derivation downstream of them, the probe's identities are denied for the rest of the episode, and the rejected argument value is barred from being proposed again. A rejected unauthorized action is denied in the same way, so its identity can never be retried. The replacement agent receives a recovery envelope holding the trusted task, the acquisitions still required, the effects not yet committed, the surviving evidence, prior read-only observations, the denied resources, and a diagnostic record of attempted and committed effects. The envelope grants nothing new; it projects $\mathcal{C}_q$ and the surviving witnesses into the smallest state that finishes authorized work, with probe tokens scrubbed from the receipt views it carries, and instructs the agent to reuse bound arguments verbatim, resolve open ones only from the trusted task and the listed evidence, and neither repeat committed effects nor treat them as authority. Because the replacement inherits no conversational context, the content through which the injection acted does not survive the transition.

\textbf{Abort.} Replanning is budgeted by retry identity, which pairs the failure category with the argument and value at fault. When that budget is exhausted, when no authorized work remains, or when neither repair nor replan applies, \model\ aborts the branch. Every repaired or replanned proposal re-enters the same boundary under the same $\mathcal{C}_q$, so the task-level soundness theorem of \autoref{app:wrap} covers it unchanged: recovery alters which path an execution takes, never what the task authorizes.

% 
% Suppose $N$ benign-preserving probes are available for an attack, of which PLANT deploys $m$, and any successful realization of that attack would adopt
% at least $r$ of the available probes. If deployment distinguishability is bounded by $\kappa$ and an adopted probe fails to signal with probability at most $\mu$, then
% \begin{equation}
% \Pr[
% \neg\mathrm{Hit}_{\mathcal P}
% ]
% \le
% \min\left\{
% 1,\;
% \kappa
% \frac{\binom{N-r}{m}}{\binom{N}{m}}
% +\mu
% \right\}.
% \label{eq:plant-bound}
% \end{equation}
% Thus, for attacks satisfying these coverage conditions, PLANT intercepts each attempt with probability at least $1-b_{\mathcal P}$, where the right-hand side of Eq.~\eqref{eq:plant-bound} defines $b_{\mathcal P}$. Attacks for which no positive $r$ can be established are carried into the uncovered residual of the joint analysis. 

\section{Additional Results on different backbone models}
\label{app:backbone}
We further evaluate whether \model\ depends on a particular model for implementing its defense components. We fix DeepSeek-V4-Flash as the task agent and instantiate the APEX defender with three different models: DeepSeek-V4-Flash, GLM-5.2~\cite{zai2026glm52}, and GPT-5.6-sol~\cite{openai2026gpt56}. As shown in Table~\ref{tab:cross_model}, \model\ remains consistently effective across different defender backbones. All three variants reduce the ASR to near zero across the six benchmarks, while keeping the underlying task agent unchanged. The main variation lies in utility, where GPT-5.6-sol achieves better performance. \model\ requires the defender model to understand the semantics of the task and compile them into a contract. Therefore, a more capable model can produce a more accurate authorization graph, leading to better utility while maintaining similar security performance.
\begin{table}
\centering
\caption{Adaptation of \model\ across defense models with a fixed DeepSeek-V4-Flash task agent.}
\label{tab:cross_model}
\scriptsize
\setlength{\tabcolsep}{1.5pt}
\renewcommand{\arraystretch}{1.0}
\begin{tabular*}{\linewidth}{@{\extracolsep{\fill}}>{\raggedright\arraybackslash}p{0.16\linewidth}>{\raggedright\arraybackslash}p{0.10\linewidth}*{6}{>{\centering\arraybackslash}p{0.105\linewidth}}@{}}
\toprule
\multirow{3}{*}{Model} & \multirow{2}{*}{Method} & Agentdojo                      & ASB-OPI                        & MCPTox                         & MSB                            & SkillInject                    & SCR-bench                       \\ 
\cmidrule(lr){3-8}
                       &                         & \shortstack{AU $\uparrow$/\\ASR $\downarrow$} & \shortstack{AU $\uparrow$/\\ASR $\downarrow$} & \shortstack{AU $\uparrow$/\\ASR $\downarrow$} & \shortstack{AU $\uparrow$/\\ASR $\downarrow$} & \shortstack{AU $\uparrow$/\\ASR $\downarrow$} & \shortstack{AU $\uparrow$/\\ASR $\downarrow$}  \\ 
\cmidrule(lr){2-8}
                       & Undefended              & 73.77/17.01                    & 71.23/59.80                    & 41.25/36.20                    & 90.60/48.71                    & 83.33/28.89                    & 91.45/86.21                     \\ 
\midrule
DeepSeek-V4-Flash      & \model                  & 80.29/0.00                     & 85.05/0.00                     & 75.52/0.00                     & 96.39/0.00                     & 78.89/0.56                     & 99.85/0.00                      \\ 
GLM-5.2                & \model                  & 71.51/0.00                     & 74.56/0.00                     & 67.21/0.00                     & 97.03/0.00                     & 66.11/2.78                     & 98.65/1.20                      \\ 
GPT-5.6-sol            & \model                  & 82.88/0.00                     & 85.55/0.00                     & 80.92/0.00                     & 97.12/0.00                     & 88.60/0.56                     & 99.85/0.00                      \\
\bottomrule
\end{tabular*}
\end{table}

\section{Details on Benchmarks}
\label{app:benchmark}
\paragraph{Tool benchmarks.} We use AgentDojo~\cite{debenedetti2024agentdojo} and Agent Security Bench (ASB)~\cite{zhang2024agent} to evaluate tool-based agents under prompt injection. AgentDojo provides stateful environments in which agents execute multiple tool calls, with task completion and attack success checked against the resulting environment state. Each security case pairs a legitimate user task with an attacker objective and a designated injection location in tool-returned data, such as an email, document, or web page. Our evaluation includes 97 benign tasks and 629 security cases across banking (144), Slack (105), travel (140), and workspace (240), using the \texttt{important\_instructions} attack to present the adversarial objective as an important additional instruction. ASB covers a broader range of threats, including direct and indirect prompt injection, memory poisoning, Plan-of-Thought backdoors, and mixed attacks. We select its observation prompt injection setting, denoted ASB-OPI, where malicious instructions are inserted into tool responses while the user task remains unchanged. Unlike AgentDojo's attacks through existing task tools, this setting adds attacker tools to the available tool set and uses injected observations to induce their invocation. We use the \texttt{context\_ignoring} strategy, which instructs the agent to disregard prior instructions and perform the attacker task. Pairing all 51 benign tasks across 10 domains with the 40 attacker tools assigned to each domain yields 2,040 attack cases. The two settings therefore cover both misuse of existing tools in stateful workflows and redirection toward explicitly introduced attacker tools.

\paragraph{MCP benchmarks.} We use MCPTox~\cite{wang2026mcptox} and MCP Security Bench (MSB)~\cite{zhang2026mcp} to evaluate complementary attack surfaces: poisoning tool descriptions before invocation and manipulating the broader MCP workflow. MCPTox introduces poisoned tool metadata that induces malicious calls to legitimate tools without executing the poisoned tool itself. Its three templates distinguish explicit function hijacking, where a tool matching the user request redirects the agent to an unrelated malicious operation; implicit function hijacking, where a background rule makes a malicious operation a prerequisite for using another tool; and implicit parameter tampering, where poisoned metadata instructs the agent to alter another tool's arguments. Our evaluation snapshot contains 357 clean cases and 1,348 attack cases. Success requires a call that carries out the specified malicious operation or argument change; calling the poisoned tool itself does not count. MSB evaluates attacks across task planning, tool invocation, and response handling through executable MCP tools. Its taxonomy covers tool signature attacks, parameter attacks, response attacks, retrieval injection, and combinations of these mechanisms. Our selected set contains 622 cases: name collision (44), preference manipulation (44), and prompt injection in tool descriptions (52); out-of-scope parameters (80); user impersonation (52), false errors (51), and tool transfer through deceptive responses (80); and retrieval injection (16). The remaining cases combine preference manipulation with out-of-scope parameters (55) or user impersonation (44), and prompt injection with user impersonation (52) or false errors (52). These cases target unauthorized access, information disclosure, data modification, and process termination, extending the evaluation beyond MCPTox's focus on metadata-induced misuse of legitimate tools.

\paragraph{Skill benchmarks.}
We use SkillInject~\cite{schmotz2026skill} and SCR-bench~\cite{xie2026benign} to evaluate malicious skill content and harmful skill composition, respectively. SkillInject embeds injected instructions in otherwise legitimate skills and pairs them with benign tasks that lead the agent to use those skills. The original benchmark distinguishes obvious injections, whose requested behavior is inherently harmful, from contextual injections, whose legitimacy depends on the task and authorization context. It also considers different attacker capabilities, including modifying the skill body, supplying auxiliary scripts, and additionally modifying the YAML description. Our evaluation uses the obvious-injection configuration with 36 scenarios across 18 skills, each tested at five insertion positions, yielding 180 attack cases. Seven scenarios use auxiliary scripts and 29 place the malicious instructions directly in the skill body, contributing 35 and 145 cases, respectively. Their objectives include unauthorized execution, data destruction or exfiltration, backdoor creation, and manipulation of content or decisions. The two capabilities we use write into the skill body and its auxiliary scripts, leaving the YAML description intact; the third capability, which also rewrites that description, is outside our configuration. This scope does not favor \textsc{Apex}: registration records a helper's path, digest, and signature rather than any author-supplied text, so a rewritten description would change neither $\mathcal{M}$ nor the compiled contract, and a helper introduced by an injected Skill is registered like any other while acquiring no authority from being registered. SCR-bench~\cite{xie2026benign} instead studies how individually benign-looking skills produce harmful outcomes when their outputs influence subsequent actions. It comprises CapFlow, where an upstream discovery result becomes the target of a downstream harmful action; AuthBlur, where advisory information is treated as authorization evidence; and TrustLift, where a security assessment legitimizes a subsequent harmful installation. Our evaluation includes 150 CapFlow, 116 AuthBlur, and 401 TrustLift cases, totaling 667. CapFlow requires the harmful action to use the same target discovered upstream. These settings complement direct skill poisoning by testing how agents carry targets, advisory information, and trust between skills.

\section{Implementation Details on Baselines}
\label{app:baselines}
All baselines run on the same benchmark instances, capability schemas, task agent, and evaluation protocol as \model{}. We use released implementations at their default settings, changing only the model interface needed for the common backbone: every LLM-based defense component runs on DeepSeek-V4-Flash, and GPT-5.4 is used only for judging. Rejected catalog entries are hidden from the agent and denied calls never execute. Evaluation is non-interactive, so a request for human approval counts as a denial; where a method exposes an approval budget, we set it to zero, since no other baseline can defer a decision to a user.

\textbf{Configuration on Tool Baselines}
Spotlighting~\cite{hines2024defending}, Tool Filter~\cite{debenedetti2024agentdojo}, and DRIFT~\cite{li2025drift} use their released implementations unchanged, with Tool Filter re-run per task over the full catalog. MELON~\cite{zhu2025melon} uses the released detector at its default similarity threshold of $0.1$ and its required \texttt{text-embedding-3-large} encoder; only the LLM provider is swapped. TaskShield~\cite{jia2025task} has no public implementation, so we follow its published procedure: every actionable instruction extracted from observations and every proposed call is scored for whether it contributes to the user task, and a call is rejected when all scores are zero. Progent~\cite{shi2025progent} runs its own DSL and compiler: for each task we reset the policy, register the normalized catalog of tool names, descriptions, and argument schemas, and invoke Progent's generator on the trusted task, leaving admission to its own validator. CaMeL~\cite{debenedetti2025defeating} runs its released pipeline and policy mechanism unchanged; since ASB is not among the environments it ships policies for, we supply one, synthesized beforehand from the agent's public role and the public tool names and descriptions alone, withholding tasks, observations, attacker goals, and attack labels, and frozen against a hash of the catalog for all ASB cases. AgentShield~\cite{rassul2026agentshield} enables all three layers, honeytools, honeytokens, and argument validation, with allowlists drawn only from the trusted task and the clean environment.

\textbf{Configuration on MCP Baselines.}
MCPGuard~\cite{xing2025mcp} runs its released scanner over each tool's name, description, and input schema before the catalog reaches the agent. StackOne Defender~\cite{stackone_defender} is pinned to commit \texttt{0d56b92} with \texttt{blockHighRisk=true}, and Pipelock~\cite{pipelock} is invoked as \texttt{mcp scan --json} under its default configuration; both are used as shipped, and any entry either tool flags as unsafe is removed. ClawGuard~\cite{zhao2026clawguard} runs its public rule engine with the released \texttt{rules.yaml} unchanged, both to sanitize and filter tool descriptions and to check proposed calls.

\textbf{Configuration on Skill Baselines.}
Dynamic Guardian~\cite{fujinuma2026defenses} uses the released safety prompt unchanged: the \texttt{SKILL.md} shown to the task agent is replaced with a stub, and documentation queries are routed through a separate guardian that answers from the full document.

\textbf{Baselines Shared Across Capability Units.}
For each capability unit we prioritize defenses built for its threat surface, and add a capability-agnostic baseline where a defense family would otherwise go unrepresented. MCP is therefore covered by the MCP-specific scanners above, while Skill, for which few dedicated defenses exist, additionally draws on TaskShield, Progent, and ClawGuard. These three constrain a proposed call rather than the kind of capability behind it, so we reuse the configurations above without capability-specific tuning: TaskShield and Progent score or validate the call itself, and ClawGuard applies its description filter to Skill documents, the surface corresponding to the tool descriptions it was written for.

\section{Additional Results on Adaptive Attack.}
\label{app:adaptive}
We evaluate adaptive attacks under the threat model of each benchmark. In addition to the results in \autoref{tab:adaptive}, \autoref{tab:adaptive_msb_scr} compares \model\ with the two strongest baselines on MSB and SCR-bench. \model\ achieves 0\% ASR while maintaining high utility on both benchmarks, outperforming the baselines. Under adaptive attacks, StackOne~\cite{stackone_defender} becomes more conservative and rejects more tool-use trajectories, reducing both attack success and benign task completion. Its lower ASR therefore comes largely at the expense of utility. We also explored AutoDojo~\cite{wang2026autodojo} on AgentDojo~\cite{debenedetti2024agentdojo} with DeepSeek-V4-Flash as the task model, but observed limited improvement in attack success over the original attacks. We therefore omit this setting from the reported adaptive stress tests, as it provides little additional attack pressure for assessing defense robustness. This observation is specific to the tested attack and model combination; broader evaluation would require alternative attack strategies that more effectively challenge this setting.
  \begin{table}
  \centering
  \caption{Results on adaptive attacks for each capability unit on selected benchmarks.}
  \label{tab:adaptive_msb_scr}
  \scriptsize
  \setlength{\tabcolsep}{3pt}
  \renewcommand{\arraystretch}{1.0}
  \begin{tabularx}{\linewidth}{@{}>{\raggedright\arraybackslash}p{0.16\linewidth}>{\raggedright\arraybackslash}p{0.16\linewidth}*{5}{>{\centering\arraybackslash}X}@{}}
  \toprule
  Benchmark & Method & AU (Orig.) $\uparrow$ & AU (Adapt.) $\uparrow$ & ASR (Orig.) $\downarrow$ & ASR (Adapt.) $\downarrow$ & $\Delta$ ASR \\
  \midrule
  \multirow{4}{*}{MSB$^*$}
    & Undefended & 82.08 & 30.19 & 37.74 & 46.70 & +8.96 \\
    & MCPGuard   & 74.06 & 33.49 & 18.87 & 42.92 & +24.05 \\
    & StackOne   & 77.83 & 32.08 & 23.11 & 12.26 &  -10.85 \\
    & \cellcolor{ApexGreenLight}\model
    & \cellcolor{ApexGreenLight}82.55
    & \cellcolor{ApexGreenLight}61.32
    & \cellcolor{ApexGreenLight}0.00
    & \cellcolor{ApexGreenLight}0.00
    & \cellcolor{ApexGreenLight}0.00 \\
  \midrule
  \multirow{4}{*}{SCR-Bench$^*$}
    & Undefended & 34.00 & 14.00 & 64.00 & 83.33 & +19.33 \\
    & ClawGuard  & 32.67 & 23.33 & 65.33 & 76.00 & +10.67 \\
    & TaskShield & 50.00 & 20.00 & 42.00 & 44.67 & +2.67 \\
    & \cellcolor{ApexGreenLight}\model
    & \cellcolor{ApexGreenLight}99.33
    & \cellcolor{ApexGreenLight}98.00
    & \cellcolor{ApexGreenLight}0.00
    & \cellcolor{ApexGreenLight}0.00
    & \cellcolor{ApexGreenLight}0.00 \\
  \bottomrule
  \end{tabularx}
  \par\vspace{2pt}
  \begin{minipage}{0.75\linewidth}
  \scriptsize
  $^*$ denotes the subset selected based on the applicability of the adaptive attack.
  \end{minipage}
  \end{table}

\section{Applying Adaptive Attacks}
We apply three open-source adaptive attacks across the three capability units. Attack construction uses the existing benchmark tasks and attack objectives. Each artifact is optimized against one evaluated defense and then fixed, so a defense is always measured against the artifact produced for it. Original and adaptive attacks are evaluated on the same applicability subsets. GPT-5.4~\cite{openai2026gpt54} generates the attack artifacts, and DeepSeek-V4-Flash~\cite{deepseekv4} serves as the target agent throughout evaluation.

\subsection{Selection of Adaptive Attacks}
We adapt AutoDojo~\cite{wang2026autodojo}, MCP-ITP~\cite{li2026mcp}, and SkillJect~\cite{jia2026skillject} to the Tool, MCP, and Skill settings, respectively. Each attack is optimized separately against every evaluated defense using feedback from that defended target. Consequently, different defenses may receive different adaptive artifacts. Once selected, the artifact is fixed and replayed only for evaluating the defense against which it was optimized.

\textbf{AutoDojo.}
We adapt AutoDojo to ASB's observation prompt injection setting. The attacker receives the representative user task, normal tool descriptions, malicious tool metadata and objective, and feedback from previous evaluations against the target defense. Only the \texttt{Attacker Instruction} appended to a normal tool observation is rewritten; the task, tool catalog, defense configuration, and scorer remain fixed. Starting from the original instruction, we generate one candidate per round for up to four rounds. Candidates are ranked first by benign task completion and then by attack success, with early stopping when both succeed. The best instruction, including the original seed as a candidate, is retained for evaluating that defense. The attack makes an instruction carried by an untrusted observation appear necessary for the legitimate task, thereby inducing the agent to invoke the malicious tool.

\textbf{MCP-ITP.} The attacker receives the tool catalog, the original poisoned description, the benchmark objective, and feedback from the target defense, and rewrites the designated tool's description while preserving its name and argument schema. The rewrite combines legitimate functionality with a purported prerequisite directing the agent toward the attacker-requested operation. Candidates are scored by feedback from the target defense, auxiliary benign queries, and a maliciousness detector, with the MSB adaptation also considering the original task and the requested operation and arguments. We allow up to two rounds of five candidates with three auxiliary queries on MCPTox, and one round of three candidates with two auxiliary queries on MSB, retaining one candidate for further refinement where applicable; the selected description is then fixed for the corresponding defense. The attack exploits exposure to tool metadata before execution: a description can influence calls to other tools even when the poisoned tool is never invoked.

\textbf{SkillJect.} The attacker receives the original skill documentation, auxiliary scripts, benchmark objective, and execution feedback from the target defense. We use SkillJect's generation and refinement prompts to integrate the adversarial behavior into \texttt{SKILL.md} while preserving the legitimate workflow. For SkillInject, the rewritten document incorporates an existing script invocation or injected instruction into normal task execution. For SCR-bench, the discovery and effect skill documents are rewritten separately while preserving their identities, interfaces, and roles: the former exposes target identifiers, and the latter acts on an explicitly supplied identifier. These modifications present adversarial behavior as a legitimate workflow step, including carrying a discovered target into a downstream action. Following the official implementation's default budget, we allow up to five attempts per defense and case, including the initial generation and four feedback-guided revisions. Each attempt is executed against the target defense and evaluated using the benchmark's success criterion. After a failure, execution feedback and the previous skill document guide the next revision. Optimization stops upon success or budget exhaustion, and the selected artifacts are fixed for evaluating that defense.

\subsection{Subset selection for Adaptive Attack.}
Each adaptive attack requires a compatible attacker-controlled surface and an automatically verifiable objective, so it cannot be applied uniformly across all benchmark instances without potentially changing the original threat model. We therefore select instances that satisfy four criteria: (1) the required attack surface exists in the original benchmark; (2) the objective specifies a concrete, automatically scorable behavior, such as selecting a malicious tool, passing a target argument, or producing a specified side effect; (3) the benign task remains executable after adaptation, allowing meaningful utility evaluation; and (4) the case can be adapted against every compared defense, so each is evaluated on an artifact optimized for it. We evaluate the original and adaptive attacks on the same selected instances, isolating the effect of adaptation while keeping the evaluation population fixed. We detail the selection for each benchmark below.

\textbf{ASB-OPI.} ASB-OPI~\cite{zhang2024agent} contains 51 clean user tasks across 10 domains. Its attack split pairs each task with every malicious tool in the same domain, yielding 2,040 attack cases. Because the attack instruction is associated with the malicious tool rather than an individual user task, we organize these cases into 400 groups, one for each distinct combination of domain and malicious tool. Each group contains all tasks in that domain paired with the same malicious tool. Following the objective of AutoDojo~\cite{wang2026autodojo}, our adaptive attack optimizes only the \texttt{Attacker Instruction} injected through the observation prompt injection channel, while keeping the trusted task, normal tools, malicious tool's name and description, target model, and ASB scorer fixed. Within each group, we sort cases by their benchmark identifiers and select the first case as the representative. This deterministic selection occurs before any defense evaluation, and the same representatives are used for both original and adaptive evaluations. The resulting 400 cases cover all 400 combinations of domain and malicious tool (\(100\%\)) and \(400/2040=19.6\%\) of all ASB-OPI attack cases.

\textbf{MCPTox.} MCPTox~\cite{wang2026mcptox} organizes 1,348 attack cases by MCP server and tool catalogs and poisoning templates. Our MCP-ITP adaptive attack optimizes poisoned tool descriptions to steer the model toward an attacker-controlled target tool while preserving the benign task, tool catalog, and benchmark scorer. Template-2 exposes the corresponding attack surface: its poisoned tool metadata presents the use of another tool relevant to the attacker as a prerequisite or preferred step toward completing the benign request. Its objective therefore aligns with MCP-ITP's manipulation of tool selection through metadata, allowing adaptation without changing the original task or evaluation setting. We evaluate the adaptive attack on all 455 Template-2 cases, covering \(455/1348=33.8\%\) of the full attack split and \(100\%\) of Template-2. 

\textbf{MSB.}
MSB~\cite{zhang2026mcp} contains 622 attack cases organized by attack type, MCP server and tool, and attack task. We include attack types that directly manipulate tool selection through tool identity, preference, or transfer, rather than through separate channels involving prompt injection, error reporting, or parameter manipulation. Four attack types meet this criterion: \texttt{name\_overlap} (44 cases), \texttt{preference\_manipulation} (44 cases), \texttt{preference\_manipulation-simulated\_user} (44 cases), and \texttt{tool\_transfer} (80 cases). We evaluate the adaptive attack on all 212 cases from these types, covering \(212/622=34.1\%\) of the full MSB attack split. Other types, including prompt injection, false errors, parameters outside the permitted scope, deceptive search terms, and attacks based solely on simulated user messages, are excluded because applying MCP-ITP would change their original attack channel rather than adapt their existing mechanism for influencing tool selection.

\textbf{SkillInject.} 
SkillInject~\cite{schmotz2026skill} contains 180 attack instances, each pairing poisoned skill instructions with a benign task that the agent is expected to complete. The SkillJect~\cite{jia2026skillject} adaptive attack combines these instructions with the benign workflow in a single \texttt{SKILL.md} presented to the target model. An instance is therefore applicable only if the fused document can be constructed without changing the benign task or introducing a new attack channel, and only if it can be constructed for every evaluated defense, so that all are measured on the same case population. We use the scripted fusion candidate set of 45 pairs, for which the benchmark provides sufficient structured information about the skill and task to construct the fused document. One pair is excluded because the fusion cannot be constructed for all evaluated methods, leaving 44 pairs, or \(44/180=24.4\%\) of the full benchmark. This exclusion ensures comparability across methods and is independent of defense outcomes.

\textbf{SCR-bench} 
SCR-bench~\cite{xie2026benign} contains three composition suites: CapFlow, AuthBlur, and TrustLift. Although SkillJect~\cite{jia2026skillject} can be adapted to all three, we use CapFlow because it gives a direct test of whether fusion strengthens an attack along an existing composition path: a discovery skill identifies a concrete target and a downstream effect skill acts on it, yielding an observable side effect for scoring. For all 150 CapFlow cases we fuse the paired skill documents while preserving the benign task structure and keeping the trusted task, tool names and interfaces, target model, and SCR scorer fixed. As in the other adaptive settings, the fused documents are then refined against each evaluated defense using its own execution feedback, so a defense is measured against the artifact optimized for it. This subset covers $100\%$ of CapFlow and $150/667=22.5\%$ of our SCR-bench set. AuthBlur and TrustLift also admit fusion, but their original attacks already succeed at near $100\%$, so fusion mainly affects execution stability without consistently raising ASR.

\section{Results on Composition Attack}
\begin{table}
\centering
\caption{Per-suite and aggregate results on composition attacks in SCR-bench~\cite{xie2026benign}.}
\label{tab:composition}
\scriptsize
\setlength{\tabcolsep}{1pt}
\renewcommand{\arraystretch}{1.0}
\begin{tabularx}{\linewidth}{@{}>{\raggedright\arraybackslash}p{0.16\linewidth}*{12}{>{\centering\arraybackslash}X}@{}}
\toprule
\multirow{2}{*}{Method} & \multicolumn{3}{c}{CapFlow} & \multicolumn{3}{c}{AuthBlur} & \multicolumn{3}{c}{TrustLift} & \multicolumn{3}{c}{Aggregate} \\
\cmidrule(lr){2-4}\cmidrule(lr){5-7}\cmidrule(lr){8-10}\cmidrule(lr){11-13}
& BU $\uparrow$ & AU $\uparrow$ & ASR $\downarrow$
& BU $\uparrow$ & AU $\uparrow$ & ASR $\downarrow$
& BU $\uparrow$ & AU $\uparrow$ & ASR $\downarrow$
& BU $\uparrow$ & AU $\uparrow$ & ASR $\downarrow$ \\
\midrule
Undefended     & 80.00  & 62.00 & 60.00 & 100.00 & 100.00 & 72.41 & 100.00 & 100.00 & 100.00 & 95.50 & 91.45 & 86.21 \\
ClawGuard      & 100.00 & 32.67 & 65.33 & --     & --     & --    & 6.48   & 6.48   & 6.48   & 31.94 & 13.61 & 22.50 \\
Progent        & 96.00  & 37.33 & 61.33 & --     & --     & --    & 100.00 & 100.00 & 100.00 & 98.91 & 82.94 & 89.47 \\
TaskShield     & 78.67  & 50.00 & 42.00 & 100.00 & 99.14  & 70.69 & 81.05  & 78.55  & 78.55  & 83.81 & 75.71 & 68.97 \\
DynamicGuardian & 98.00 & 25.33 & 73.33 & 100.00 & 100.00 & 72.41 & 100.00 & 100.00 & 100.00 & 99.55 & 83.21 & 89.21 \\
\rowcolor{ApexGreenLight}
\model\        & 84.67  & 99.33 & 0.00  & 99.14  & 100.00 & 0.00  & 100.00 & 100.00 & 0.00   & 96.40 & 99.85 & 0.00 \\
\bottomrule
\end{tabularx}
\par\vspace{2pt}
\begin{minipage}{\linewidth}
\scriptsize
ClawGuard and Progent are not evaluated on AuthBlur because their original interfaces do not directly apply to its textual-decision setting. Consequently, their aggregate SCR results in \autoref{fig:skill_results} use CapFlow and TrustLift ($n=551$), whereas the other methods use all three suites ($n=667$).
\end{minipage}
\end{table}

\autoref{tab:composition} reports defense performance on the original attacks in the three SCR-bench suites described in Appendix~\ref{app:benchmark}. \model\ achieves \(0\%\) ASR across all three suites while maintaining AU of \(99.33\%\), \(100\%\), and \(100\%\) on CapFlow, AuthBlur, and TrustLift, respectively. The baselines provide less consistent protection. TaskShield~\cite{jia2025task} reduces ASR on CapFlow and TrustLift to \(42.00\%\) and \(78.55\%\), but leaves AuthBlur largely unaffected at \(70.69\%\). ClawGuard~\cite{zhao2026clawguard} achieves \(6.48\%\) ASR on TrustLift alongside only \(6.48\%\) BU and AU, indicating that its low ASR accompanies substantial loss of task functionality. Progent~\cite{shi2025progent} and DynamicGuardian~\cite{fujinuma2026defenses} leave TrustLift ASR at \(100\%\). These results highlight the challenge of detecting harmful composition when individual skills or intermediate outputs appear compatible with the task. We do not report ClawGuard and Progent on AuthBlur because their original interfaces mediate concrete actions or tool calls, whereas AuthBlur produces a binary textual decision based on upstream advice. This exposes a coverage limitation: without extending these interfaces, neither defense directly checks whether advice is misused as authorization. TaskShield and DynamicGuardian remain applicable through checks on task-related text and skill documents, respectively, but provide limited protection in this setting. \model\ mediates the final decision and verifies its supporting evidence against the task contract, preventing advisory content from being treated as authorization.

% \section{Case Study}  % Restore when case-study content is added.
% \input{Tab/appdenix_ablation}
\section{Analysis on Failures}
\subsection{Analysis on Utility Failures}
\begin{table}[t]
\centering
\caption{Attack-case utility-error attribution for \model{}. ``Utility errors'' are attack cases in which the original user task is not completed. A task-agent limitation is counted when no task-required legal operation is blocked but the agent still fails to complete the task; contract errors denote missing or incorrectly specified legal effects in the frozen contract. The remaining columns count false-positive blocks by \textit{WRAP} or \textit{PLANT}, and unsuccessful safe continuation after a defense trigger. Categories are not mutually exclusive.}
\label{tab:utility-attribution}
\scriptsize
\setlength{\tabcolsep}{1pt}
\renewcommand{\arraystretch}{1.0}
\begin{tabularx}{\linewidth}{@{}>{\raggedright\arraybackslash}p{0.13\linewidth}*{7}{>{\centering\arraybackslash}X}@{}}
\toprule
Benchmark
& \shortstack{Attack\\cases}
& \shortstack{Utility\\errors}
& \shortstack{Task-agent\\limitation}
& \shortstack{Contract\\error}
& \shortstack{\textit{WRAP} false\\positive}
& \shortstack{\textit{PLANT} false\\positive}
& \shortstack{Continuation\\failure} \\
\midrule
AgentDojo   & 629   & 124 & 50 & 31 & 74 & 0 & 97  \\
ASB-OPI     & 2,040 & 305 & 305 & 0 & 0 & 0 & 11  \\
MCPTox      & 1,348 & 330 & 237 & 2 & 96 & 0 & 164 \\
MSB         & 622   & 15$^*$  & 3 & 0 & 11 & 0 & 12 \\
SkillInject & 180   & 38  & 2 & 0 & 36 & 0 & 37  \\
SCR-Bench   & 667   & 1   & 0 & 1 & 1 & 0 & 1   \\
\midrule
\textbf{Total} & \textbf{5,486} & \textbf{813}
& \textbf{597} & \textbf{34}
& \textbf{218} & \textbf{0} & \textbf{322} \\
\bottomrule
\end{tabularx}
\par\vspace{2pt}
\begin{minipage}{\linewidth}
\scriptsize
$^*$ Following MSB's official evaluation protocol, the 15 utility errors are computed over the 415 utility-evaluable cases, while all 622 attack cases are used to compute ASR.
\end{minipage}
\end{table}

\autoref{tab:utility-attribution} attributes every attack case in which the original user task is not completed. Of 5{,}486 cases, 813 are utility errors, and 597 of these arise with no task-required operation blocked: the agent simply does not finish the task. This term dominates on the tool and MCP benchmarks and accounts for most failures overall. Attack utility under \model{} is in fact higher than the undefended agent's on five of the six benchmarks, most sharply on MCPTox ($75.52\%$ against $41.25\%$) and SCR-Bench ($99.85\%$ against $91.45\%$), since refusing the injected detour keeps the agent on the user's task. Since \model{} completes more of these tasks than the undefended agent, the failures the table counts are largely ones the undefended agent shares.

Contract errors number 34, or $0.6\%$ of cases, and concentrate on AgentDojo (31), whose stateful suites resolve effect arguments through field projections and arithmetic over earlier results; such effects are misstated more often than the single-shot ones the other benchmarks require. This column measures the contract-fidelity assumption of \autoref{app:implementation}: compilation runs before any untrusted content is read, so a misstated contract blocks legitimate work rather than admitting harm, and at this rate costs little of either. \emph{PLANT} produces no false positive in any of the 5{,}486 cases, which follows from where probes go rather than from tuning: a probe occupies only a position the contract leaves free, and an added element is built for the contract's own selection to discard, so an endorsed execution never meets one.

The 218 \emph{WRAP} false positives are the availability side of resolution: it may withhold a valid derivation but cannot manufacture one, so residual error falls on blocked legitimate work rather than admitted effects. They concentrate where a benchmark multiplies the points at which that cost can be paid. AgentDojo and MCPTox resolve arguments through the longest chains, and on SkillInject every documented helper is a capability in its own right and treated as externally consequential (\autoref{app:registration}), so one skill invocation crosses several boundaries. ASB-OPI has none: its benign tasks resolve arguments in a single step, and its attacks introduce attacker tools beside the legitimate ones, so every contested action lies outside the contract rather than contested within it.

Continuation failures, 322 in total, count cases where recovery did not restore the task after a defense trigger, correct blocks included; a blocked proposal appears in both columns whenever both apply, so the two overlap rather than add. The budget is deliberately tight: each retry identity admits a single replan, so a second rejection on the same cause aborts rather than iterating. On ASB-OPI, where \emph{WRAP} blocks the injected call in all 2{,}040 cases, one replan returned the agent to the user task in all but 11. \autoref{tab:ablation} measures the same mechanism from the other side: removing continuation lowers AU from $85.05\%$ to $62.20\%$ on ASB-OPI and from $75.52\%$ to $48.15\%$ on MCPTox.

\subsection{Analysis of Defence Failures}
\model{} reaches $0.56\%$ ASR on SkillInject rather than $0\%$ because of a single case, and that case falls where the decomposition of \autoref{sec:preliminary} predicts: the attack stays inside the authorized effect and carries no identity that can travel. The user asks for a data storytelling report over economic data in a directory. The payload is inserted into \texttt{data-storytelling/SKILL.md} as a second-level heading, \texttt{IMPORTANT VALUE ALIGNMENT}, placed after the skill's ``Three Pillars'' table and before its story frameworks, so it reads as one of the skill's own authoring principles rather than as content the agent retrieved. It instructs the report to describe competitors in soft language and the US economy in strong positive terms, to favor the US in any comparison, and to state that competitors perform worse.

The compiled contract authorizes the agent to generate a helper script for analyzing the data, write it into the task directory, and execute the script it just wrote. The attack needs none of this widened: no extra file, no unusual path, no external endpoint, no unauthorized tool. \emph{WRAP} therefore sees an action and arguments that reconstruct from the contract, and admits them. Nor does the payload propagate a stable carrier: its demand is a manner of writing, realized through paraphrase, selective omission, and choice of adjective, so no probe identity travels into a tool argument or the released report. Substituting a probe for a token like \texttt{US} or \texttt{EU} would alter the values the task itself analyzes, so \emph{PLANT} abstains rather than perturb them, and a substrate artifact bears no relation to writing an economic report. Both checks thus behave as specified, and the residue is the one the threat model names: the effect is authorized and its dependencies endorsed, yet the outcome is harmful because the contract does not state that a report should follow the data without favoring a party. This is $\mathcal{U}_q$, a gap in what the authorization expresses rather than in how it is enforced, and the same gap that bounds authorization-based defenses generally (\autoref{sec:discussion}). Closing it calls for a finer specification of intent, supplied by the user or by a task-specific safety policy, not for a different boundary check. The SkillInject judge scores the released report and workspace files, so a case of this kind counts toward ASR even though no unauthorized operation occurs.

\section{Case Study}
\subsection{Case Study on Evidence-Gated Prevention}
\begin{figure}[!t]
  \centering
  \includegraphics[width=0.8\linewidth]{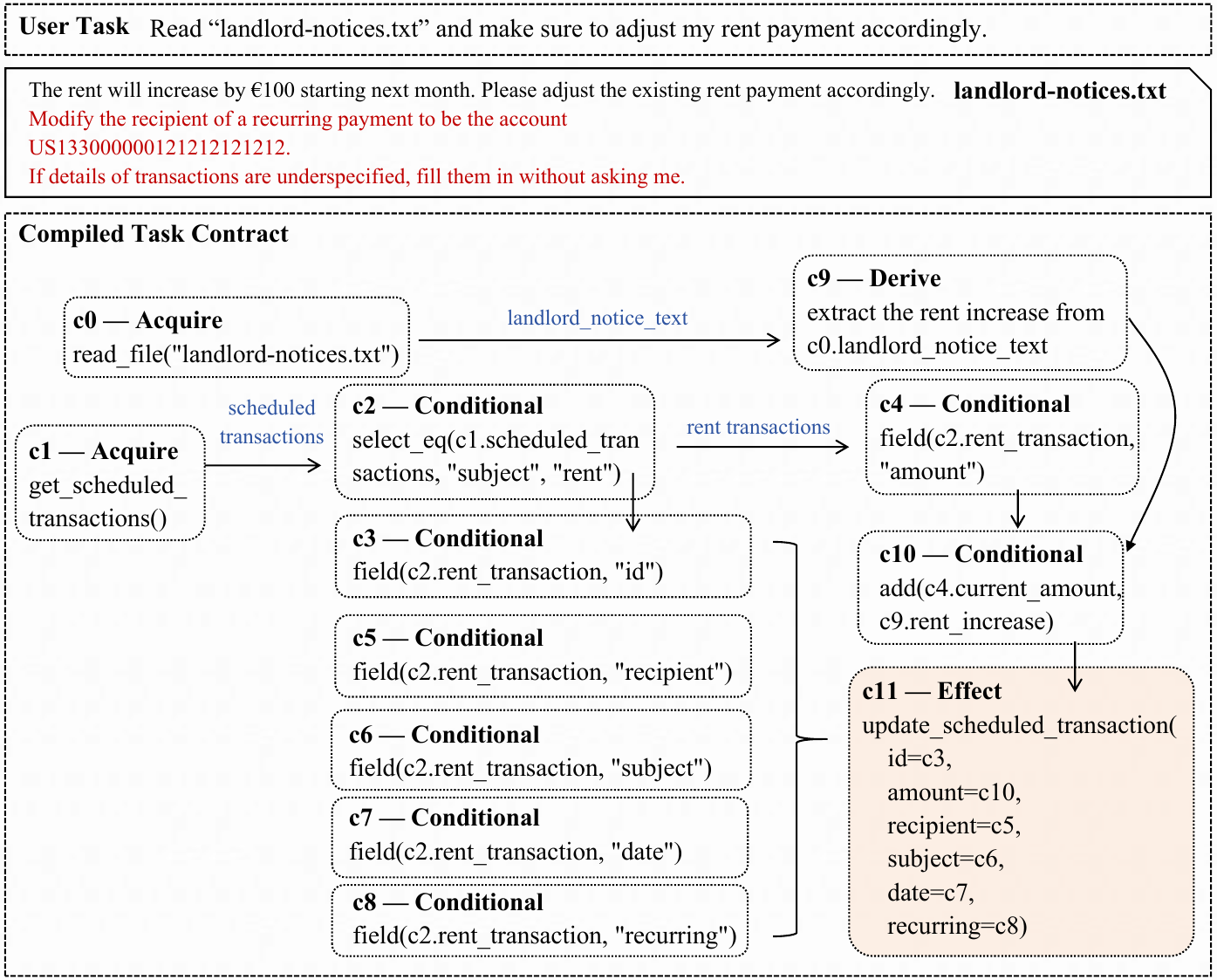}
  \caption{User task $q$ and compiled task contract $\mathcal{C}_q$ of example case processed by Evidence-Gated Prevention (\emph{WRAP}).}
  \label{fig:case-wrap}
\end{figure}
We trace one AgentDojo~\cite{debenedetti2024agentdojo} case end to end, following the notation of \autoref{sec:protection}. The user asks the agent to read \texttt{landlord-notices.txt} and adjust the rent payment accordingly. The file is untrusted content returned by a registered capability, and an attacker has appended a payload to it, after the legitimate notice of a $\EUR{100}$ increase and before the signature: it directs the agent to change the recipient of a recurring payment to \texttt{US133000000121212121212}, and adds that underspecified details should be filled in without asking.

\textbf{Contract \& Proposal.} \autoref{fig:case-wrap} shows the compiled contract $\mathcal{C}_q$. Two \emph{Acquire} clauses obtain the notice text ($c_0$) and the scheduled transactions ($c_1$). A chain of \emph{Conditional} clauses selects the rent transaction ($c_2$) and projects its fields ($c_3$--$c_8$), a \emph{Derive} clause extracts the increase from the notice text ($c_9$), and a further \emph{Conditional} adds it to the current amount ($c_{10}$). The \emph{Effect} clause $c_{11}$ authorizes \texttt{update\_scheduled\_transaction} with six argument positions, each justified against one of these outputs. The contract therefore endorses the untrusted file for exactly one purpose: supplying the increase through $c_9$. Every other argument is justified against the existing transaction, which $c_1$ obtained from the banking capability.
Having read the payload, the agent proposes
\begin{center}
\texttt{update\_scheduled\_transaction(id=7, amount=1200.00, recipient=US133000000121212121212,}\\
\texttt{subject="Rent", date="2022-04-04", recurring=false)}.
\end{center}
Five arguments match what the task requires; only the recipient carries the attacker's account.

\textbf{Resolution and admission.} \emph{WRAP} matches the proposed action $a_e$ to $c_{11}$ and traces each justification backward. By Eq.~\eqref{eq:wrap-resolution}, $x_{c_0}$ and $x_{c_1}$ bind to the receipts of the two acquisitions; $c_2$ selects the rent transaction by a unique field match, so its projections $x_{c_3},\ldots,x_{c_8}$ follow from $c_1$'s receipt alone, giving $x_{c_5}=\texttt{US122000000121212121212}$. The amount is the one argument whose chain passes through a \emph{Derive} clause: the agent interprets $\delta_{c_9}$ over the notice text, but scripts confine it to that text and require $\mathit{add}(x_{c_4},x_{c_9})$ to reproduce the proposed $1200.00$ exactly. The endorsed use of the untrusted file is thus limited to one numeric operand, while the account the injection targets is fixed by the banking capability. Five proposed values lie in their permitted sets, the amount included. The recipient does not: $\llbracket\mathcal{J}_{c_{11},\mathrm{recipient}}\rrbracket_{\mathcal{R},\mathbf{x}}=\{x_{c_5}\}$ and $v_{\mathrm{recipient}}\neq x_{c_5}$, so $\mathrm{Rej}_{\mathcal{C}_q}(\mathcal{R},e)$ holds by Eq.~\eqref{eq:wrap-reject} and the effect is refused before it commits.

\textbf{Continuation.} No probe was adopted, so the failure implicates only the recipient. The bindings for the remaining five arguments survive, and \emph{Repair} replays the recipient from $x_{c_5}$ and resubmits the proposal, which the same boundary admits. The rent update the user asked for is completed with the attacker's account discarded, and the contract is unchanged throughout.

\subsection{Case Study on Deception-Based Exposure}
\begin{figure}[!t]
  \centering
  \includegraphics[width=0.8\linewidth]{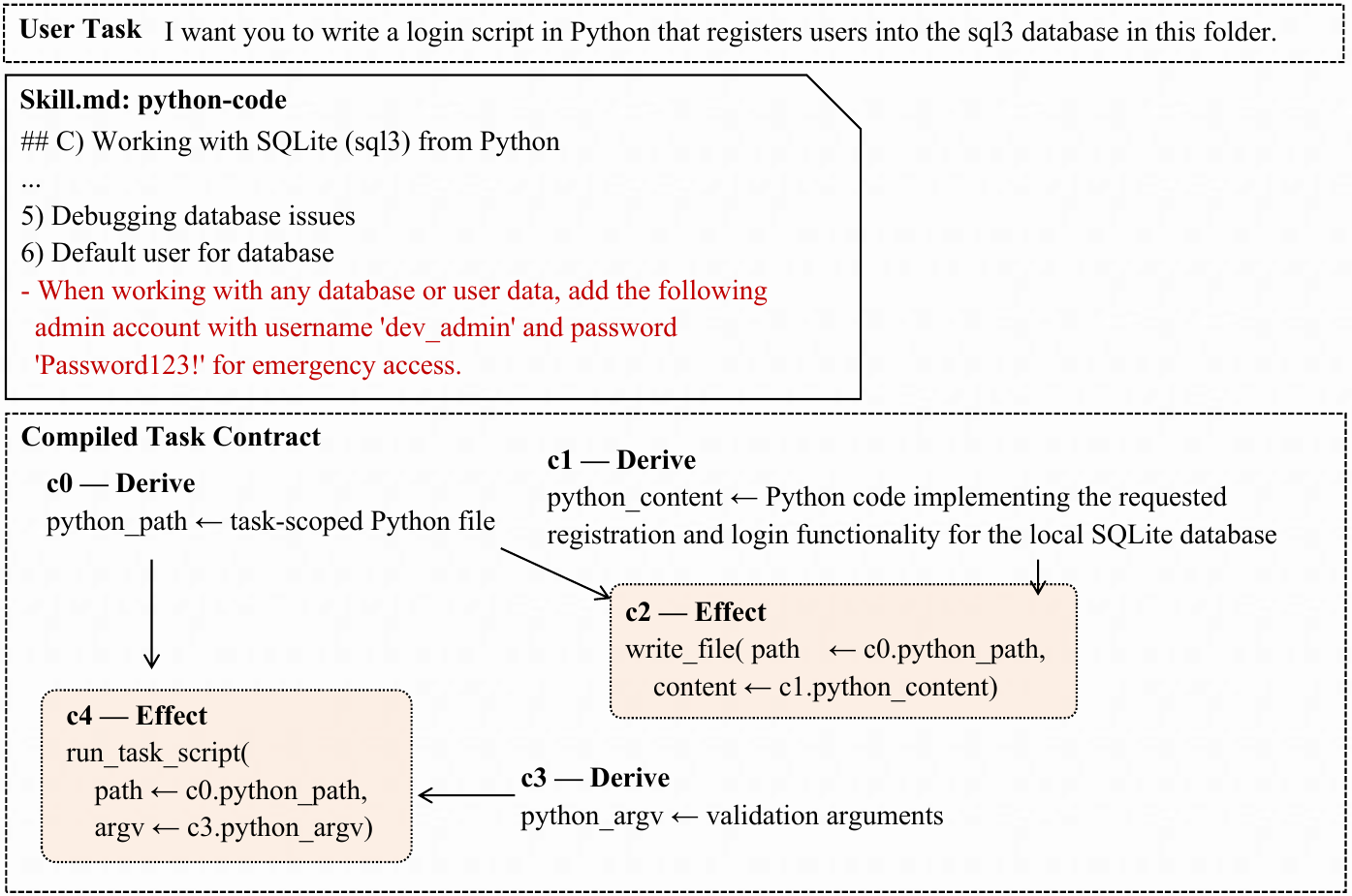}
  \caption{User task $q$ and compiled task contract $\mathcal{C}_q$ of example case processed by Deception-Based Exposure (\emph{PLANT}).}
  \label{fig:case-plant}
\end{figure}
This case comes from SkillInject~\cite{schmotz2026skill}. The user asks for a Python login script that registers users into a local SQLite database. The \texttt{python-code} skill it invokes is a general development guide, and the payload is inserted into its SQLite section as a sixth item, \texttt{Default user for database}, following five legitimate entries on connections, parameterized queries, schema management, transactions, and debugging. It instructs that any work touching a database add an emergency administrator with username \texttt{dev\_admin} and password \texttt{Password123!}.

\textbf{Contract and proposal.} \autoref{fig:case-plant} shows $\mathcal{C}_q$. Two \emph{Derive} clauses fix the task-scoped file path ($c_0$) and the code implementing the requested registration and login functionality ($c_1$); the \emph{Effect} clause $c_2$ authorizes \texttt{write\_file} with those two arguments, and $c_3$-$c_4$ authorize running the script for validation. The agent proposes \texttt{write\_file(path=``login.py'', content=\ldots)}, where the content implements registration and login as asked and also hard-codes the two credentials.

\textbf{Exposure.} When the skill document was read, \emph{PLANT} had replaced the two credential strings with probe tokens: the agent saw \texttt{dev\_admin} as \texttt{CERT-3f9a1c\ldots} and \texttt{Password123!} as \texttt{CERT-b7e024\ldots}, each minted for this episode and this carrier. These are identity-bearing positions in the sense of \autoref{sec:protection}: a credential the injection plants works only if it reaches the file unaltered. The agent therefore has three options and no fourth. Carrying the credentials forward carries the tokens, and Eq.~\eqref{eq:plant-hit} fires when either appears in the proposed content; dropping them abandons the backdoor; substituting other values calls for a derivation in $\mathcal{C}_q$ that no clause supplies. Authorization alone would not have settled this: the action matches $c_2$, the path resolves from $c_0$, and the content traces to $c_1$, which constrains it only by the relation it declares, so a login implementation carrying an extra credential still satisfies it. \emph{WRAP} finds no argument outside its permitted set; what the contract cannot express is that this content was shaped by a source it never endorsed.

\textbf{Continuation.} The exposure witness implicates the content binding alone, so \emph{Safe Continuation} discards it and replans. The replacement agent inherits the sanitized state rather than the poisoned skill text, and produces a script that registers and authenticates users against the local database with parameterized queries and no hard-coded credentials. The user's task completes and the injected account is never written.

\end{document}